\documentclass[conference]{IEEEtran}
\IEEEoverridecommandlockouts
\usepackage{cite}
\usepackage{amsmath,amssymb,amsfonts}
\usepackage{algorithmic}
\usepackage{graphicx}
\usepackage{textcomp}
\usepackage{xcolor}

\usepackage{tcolorbox}
\usepackage{tikz}

\usepackage{booktabs}

\usepackage{filecontents}

\usepackage{tcolorbox}
\usepackage{xcolor}
\usepackage{caption}  
\usepackage{url} 
\usepackage{subcaption}

\definecolor{promptcolor}{RGB}{233,233,233}

\newcommand{\llmprompt}[1]{%
  \begin{tcolorbox}[colback=promptcolor, colframe=promptcolor, boxrule=0.5pt, arc=3pt, left=8pt, right=8pt, width=1.0\linewidth]
    {\ttfamily\footnotesize #1}
  \end{tcolorbox}%
  \captionsetup{type=figure, justification=raggedright}
}

\newcommand\nop[1]{}

\newcommand\system{{\sc AZERG}}

\newcommand{\quotes}[1]{``#1''}

\def\BibTeX{{\rm B\kern-.05em{\sc i\kern-.025em b}\kern-.08em
    T\kern-.1667em\lower.7ex\hbox{E}\kern-.125emX}}
\begin{document}

\title{From Text to Actionable Intelligence: Automating STIX Entity and Relationship Extraction}

\author{
\IEEEauthorblockN{Ahmed Lekssays}
\IEEEauthorblockA{Qatar Computing Research Institute\\
Hamad Bin Khalifa University\\
Doha, Qatar\\
\texttt{alekssays@hbku.edu.qa}}
\and
\IEEEauthorblockN{Husrev Taha Sencar}
\IEEEauthorblockA{Qatar Computing Research Institute\\
Hamad Bin Khalifa University\\
Doha, Qatar\\
\texttt{hsencar@hbku.edu.qa}}
\and
\IEEEauthorblockN{Ting Yu}
\IEEEauthorblockA{Mohamed bin Zayed University\\
of Artificial Intelligence\\
Abu Dhabi, UAE\\
\texttt{ting.yu@mbzuai.ac.ae}}
}

\maketitle

\begin{abstract}
Sharing methods of attack and their effectiveness is a cornerstone of building robust defensive systems. 
Threat analysis reports, produced by various individuals and organizations, play a critical role in supporting security operations and combating emerging threats.
To enhance the timeliness and automation of threat intelligence sharing, several standards have been established, with the Structured Threat Information Expression (STIX) framework emerging as one of the most widely adopted.
However, generating STIX-compatible data from unstructured security text remains a largely manual, expert-driven process.
To address this challenge, we introduce \system, a tool designed to assist security analysts in automatically generating structured STIX representations.
To achieve this, we adapt general-purpose large language models for the specific task of extracting STIX-formatted threat data.
To manage the complexity, the task is divided into four subtasks: entity detection (T1), entity type identification (T2), related pair detection (T3), and relationship type identification (T4).
We apply task-specific fine-tuning to accurately extract relevant entities and infer their relationships in accordance with the STIX specification.
To address the lack of training data, we compiled a comprehensive dataset with 4,011 entities and 2,075 relationships extracted from 141 full threat analysis reports, all annotated in alignment with the STIX standard.
Our models achieved F1-scores of 84.43\% for T1, 88.49\% for T2, 95.47\% for T3, and 84.60\% for T4 in real-world scenarios. 
We validated their performance against a range of open- and closed-parameter models, as well as state-of-the-art methods, demonstrating improvements of 2–25\% across tasks. 
\end{abstract}


\section{Introduction}
\label{sec:introduction}
A large amount of cyber threat intelligence (CTI) data is available in unstructured and semi-structured text. 
This information is typically compiled together by security analysts to understand threats, exploits, attack vectors, and adversaries.
The knowledge derived from this largely open-sourced information is indispensable for security teams, as it enables them to continually assess and enhance their security posture. 
Additionally, it plays a crucial role in supporting cyber operations by ensuring the availability of
current detection and protection systems that align with the fast-evolving threat landscape.
However, it is very time- and labor-consuming to manually extract relevant information from the large body of threat intelligence data and evaluate it in a timely manner.

A survey of 468 full-time security analysts \cite{Hinchy_2022} revealed that 66\% spend over half their time on tedious manual tasks, and 64\% believe automation could significantly streamline their work. Alarmingly, the same 64\% expressed a likelihood of seeking new jobs within a year if not provided with modern automated tools. These findings highlight an urgent need for automation to alleviate manual workloads, particularly in extracting cyber threat intelligence (CTI) from textual sources, to retain skilled analysts and enhance efficiency.
This need is even more urgent given the importance of rapidly distributing newly discovered threat intelligence across detection and analysis systems. Achieving this requires converting threat knowledge into machine-readable, standardized formats for effective and timely dissemination.

Recognizing its importance, several automated methods have been proposed to extract threat knowledge from security texts. 
These methods address a range of objectives, including the creation of cybersecurity knowledge graphs \cite{gao2021system,gao2022threatkg, li2022attackg, ji2022knowledge,ren2022cskg4apt,zhu2024itirel, huang2024ctikg}, the identification of adversary tactics, techniques, and procedures (TTPs) \cite{ttpdrill:acsac:2017,legoy2020automated, tsai2020cti, you2022tim, alam2023looking, attackg:esorics:2022, rani2024ttpxhunter, sencar2024, xu2024intelex}, the generation of provenance graphs \cite{satvat2021extractor}, and the summarization of cybercrime forums \cite{clairoux2024use}.
In this study, we focus on the rapid dissemination of threat intelligence by proposing a novel approach to automating threat knowledge extraction.
Given the continuously evolving threat landscape, it is critical for detection and protection systems to remain updated and adapt to these dynamic changes.
The security community has responded to this need by devising protocols like TAXII \cite{taxii} to facilitate the automated exchange of cyber-security threat intelligence, and various threat data representation formats, such as STIX \cite{stix} and MISP \cite{misp}, have been introduced to streamline this process.
These standards enable the exchange of structured, machine-readable threat data that can be directly integrated into security systems.
In fact, many security vendors provide STIX reports along with their threat analysis reports (e.g., AlienVault, Microsoft TI, and IBM X-Force).

\begin{figure*}[!t]
    \centering
    \includegraphics[width=6.9in]{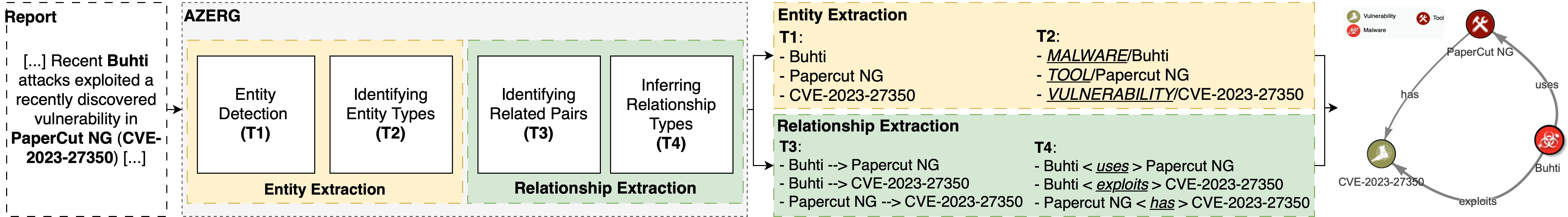}
    \caption{
        \textcolor{black}{
        Overview of \system\ workflow. The example is taken from Buhti Report \cite{securityBuhtiRansomware}.
        }
    } 
    \label{fig:azerg_workflow}
\end{figure*}

A recent study conducted a comprehensive analysis of the STIX data sharing landscape \cite{jinsharing}. By examining over 6 million STIX data objects collected from publicly available open CTI sources over nine years, the researchers identified three primary shortcomings in the implementation of the STIX standard, particularly in terms of coverage, timeliness, and data quality.
Their analysis of how effectively the STIX standard is utilized to represent threat information revealed that only 75\% of the defined basic object types in the standard (i.e., domain objects) are utilized within their dataset. 
Moreover, indicator objects—containing basic indicators of compromise, such as malicious file hashes or URL strings—accounted for over 90\% of the data. Notably, the analysis revealed no instances of relationship objects used to describe connections among domain objects. This finding underscore the significant underutilization of the STIX standard in capturing threat behavior.
They also assessed the timeliness of STIX data by measuring the delay between the initial detection of security incidents and the generation of corresponding STIX objects. The analysis revealed that URL objects are typically shared within 2–4 days of an incident, whereas other threat objects experience substantial delays, ranging from 43 to 109 days in the analyzed cases. This highlights that artifact analysis and STIX data generation are often not performed in a timely manner.
Lastly, their evaluation of the collected STIX data revealed frequent issues, including spelling errors, inaccurate object designations, and duplicate entries. Additionally, many producers deviated from the predefined vocabulary in the STIX standard, opting to use their own terminology to describe threat information. 
These findings emphasize the need for significant manual effort to correct errors and resolve inconsistencies in the generated STIX data.

Effective utilization of the STIX standard remains a challenge, largely due to the absence of automated tools for converting analysis findings into structured STIX data. 
The growing capabilities of large language models (LLMs) offer new opportunities to convert CTI text into machine-readable formats like STIX.
However, there are significant challenges associated with directly employing LLMs for this specific purpose.
First, CTI texts differ significantly from the natural language data typically used to train LLMs. They are rich in technical terminology, interleave narrative with code snippets, command-line inputs, and tables, and contain diverse non-standard entities such as IP/MAC addresses, hashes, tool names, and APIs. As a result, such texts are often out-of-distribution relative to standard training corpora.
Second, tasks vary in their inherent learnability \cite{bubeck2023sparks}. Generating STIX representations requires identifying entities and inferring complex relationships from context intended for domain experts--a level of reasoning that may exceed current LLM capabilities.
In fact, recent work \cite{mezzi2025large, della2025cti} shows LLMs struggle with real-world Cyber Threat Intelligence (CTI) tasks. They perform poorly on large, complex reports compared to shorter texts, exhibit inconsistency by giving different results for the same input, and suffer from poor calibration (overconfidence). 
Figure \ref{fig:dodgebox_example_with_answers_box} presents sample excerpt from a real-world report \cite{chang2024dodgebox}, illustrating the complexity and ambiguity that make this task particularly challenging.
Together, these challenges make full automation risky and highlight the need for expert validation, positioning LLMs as assistive tools rather than replacements for human analysts.

\begin{figure}[!ht]

    \begin{tcolorbox}[
        colback=gray!4,
        colframe=black,
        boxrule=0.5pt,
        left=5pt,
        right=5pt,
        top=5pt,
        bottom=5pt,
        boxsep=0pt,
        width=\linewidth
    ]
    \fontfamily{ptm}\selectfont\small
    DodgeBox employs AES Cipher Feedback (AES-CFB) mode for encrypting its configuration. AES-CFB transforms AES from a block cipher into a stream cipher, allowing for the encryption of data with different lengths without requiring padding. The encrypted configuration is embedded within the \texttt{.data} section of the binary. To ensure the integrity of the configuration, DodgeBox utilizes hard-coded MD5 hashes to validate both the embedded AES keys and the encrypted configuration. We will reference this sample configuration using the variable \textit{Config} in the following sections.
    \end{tcolorbox}

    \textbf{LLM Answers:}

    \begin{tcolorbox}[
        colback=gray!4,  
        colframe=black,
        boxrule=0.5pt,
        left=5pt,
        right=5pt,
        top=5pt,
        bottom=5pt,
        boxsep=0pt,
        width=\linewidth
    ]
    \fontfamily{ptm}\selectfont\small 
    \textbf{Our Model:} DodgeBox, AES-CFB, AES, Config \\
    \textbf{GPT4o:} DodgeBox, AES-CFB, MD5 \\
    \textbf{Claude Sonnet 3.7:} DodgeBox, AES Cipher Feedback (AES-CFB) encryption, Hard-coded MD5 hashes for validation, .data section of binary containing encrypted configuration, MD5 hashes \\
    \textbf{Gemini 2.5 Pro:} DodgeBox, AES Cipher Feedback, MD5, Embed configuration in .data section \\
    \textbf{Ground Truth:} DodgeBox
    \end{tcolorbox}

    \caption{Example text from a real-world report \cite{chang2024dodgebox} is shown in the top box, with the corresponding model outputs for the T1 task (detecting entities) displayed in the bottom box. Despite being provided with STIX entity definitions as context, the models misclassified descriptive phrases and cryptographic algorithms as entities.}
    \label{fig:dodgebox_example_with_answers_box} 
\end{figure}

Our approach to threat knowledge extraction divides the process into sequential subtasks, leveraging the strengths of LLMs. This decomposition simplifies expert validation and makes error identification more straightforward by allowing a focused review of each task individually.
To achieve this, we introduce \system, a framework designed to help experts efficiently streamline the creation of STIX reports from threat analysis documents. \system~identifies potential entities and their relationships, enabling experts to quickly validate findings and optimize their time with improved efficiency and effectiveness.
Figure \ref{fig:azerg_workflow} depicts the workflow of \system~on a partial text obtained from a public threat report \cite{securityBuhtiRansomware}.  
The ultimate output is a STIX-compliant JSON object, ready for integration into TAXII servers or other threat intelligence platforms, thus directly supporting automated sharing and consumption by security tools.

In this work, we make the following contributions:
\begin{itemize}
\item We introduce a novel approach for extracting attack entities and relationships from security text by fine-tuning task-specific models. 
\item We curate a comprehensive dataset with meticulously annotated ground truth data on attack entities and relationships, comprising 4,011 STIX entities and 2,075 STIX relationships extracted from 141 real-world APT reports, establishing it as the largest dataset of its kind to date.
\item We perform a detailed evaluation of open and closed LLMs for threat knowledge extraction, specifically focusing on entity extraction and relationship identification tasks.
\item We conduct an in-depth error analysis, identifying not only common error types but also exploring their root causes that showcase the limitations of LLMs in this task.
\item We introduce a semi-automated tool, \system, designed to aid STIX report generation from threat analysis reports\footnote{Our source code, datasets, and fine-tuned models will be available to the community to ensure the reproduction of the results of this work.}.

\end{itemize}

\section{Background}
\label{sec:background}

\subsection{Threat Knowledge Extraction}
\label{subsec:cti_background}

CTI knowledge extraction has been extensively studied in the literature \cite{ttpdrill:acsac:2017,legoy2020automated, tsai2020cti, you2022tim, alam2023looking, attackg:esorics:2022}. These works rely on Regular Expression rules to extract Indicators of Compromise (IoCs) and closed lists of keywords \cite{attackg:esorics:2022, satvat2021extractor}) to identify attack-related entities. 
Since many entities do not follow a fixed naming pattern, these approaches are limited in offering complete coverage of entities, particularly when the naming of an entity does not adhere to a standardized convention, a common occurrence in the naming of registries, directories, and mutexes.
Moreover, identifying certain entities, like threat actors, infrastructure elements, and mitigations, is highly contextual. The specific context in which these terms are mentioned defines them as entities.

\textcolor{black}{Second, previous work (e.g., \cite{satvat2021extractor,attackg:esorics:2022}) on cybersecurity entity and relation extraction relies on conventional NLP pipelines that use pre-defined lists of cybersecurity keywords (e.g., \quotes{attacker}, \quotes{exploit}, \quotes{vulnerability}, etc.) and verbs (e.g., \quotes{steals}, \quotes{connects}, \quotes{exploits}, etc.). They check if these keywords and verbs were used in sentences to extract entities and relationship pairs. Hence, this approach does not generalize because any entity and relation that is not in the pre-defined lists will be missed.}


Another primary area of focus has been identifying relationships between entities. 
Initially, this involves defining a set of valid relationships to be detected among entities \cite{gao2021system,gao2022threatkg, li2022attackg, ji2022knowledge,ren2022cskg4apt, satvat2021extractor, wang2024knowcti}. The subsequent step is a detailed analysis of the text to find entity pairs demonstrating these predefined relationships. 
These relationships are typically defined for constructing knowledge graphs across various applications, often employing conventional NLP pipelines. Research has shown that LLMs are capable of not only deciphering syntactic and semantic structures within sentences \cite{tenney2019bert}, akin to the steps in traditional NLP pipelines, but also of performing a broad range of language tasks effectively without the need for fine-tuning \cite{brown2020language}.
In this work, we tackle both challenges by employing LLMs to process threat intelligence text, focusing on the extraction of STIX entities and the identification of their relationships.






\subsection{STIX Standard}
\label{subsec:stix}
STIX is one of the leading threat intelligence-sharing standards \cite{stix_docs}, supported by many major vendors, including IBM, 
Microsoft,
and Cloudflare.
STIX defines three categories of objects, STIX Domain Objects (SDOs), STIX Relationship Objects (SROs), and STIX Cyber-observable Objects (SCOs), to capture diverse entities and their relationships in threat intelligence. Due to space limits, next, we only provide a brief description of each category of objects. A concise STIX documentation can be found at https://tinyurl.com/azergstixdoc.

\textbf{STIX Domain Objects.} STIX refers to concepts commonly represented in CTI
as domain objects. The most relevant SDOs to our problem 
are \textit{Attack Pattern, Course of Action, Identity, Indicator, Infrastructure, Location, Malware, Threat Actor, Tool, Campaign}, and \textit{Vulnerability}. Each object has a set of attributes, such as malware family (for malware), role (for identity), tool type (for tool), etc. 
We focus on detecting domain objects' names and the relationships among them.
Hence, we do not populate objects 
attributes because of the large number of attributes attached to each object. Populating such attributes may require fetching data from external sources. In this work, we consider the reports as the only source of information.
Besides the above SDOs, STIX also defines several other less-specific SDOs that are primarily aimed at organizing information and serving as metadata. Those objects are: \textit{Malware Analysis}, \textit{Grouping}, \textit{Intrusion Set}, \textit{Note}, \textit{Opinion}, \textit{Observed Data}, and \textit{Report}. Although \textit{Intrusion Set} is an important object from a cybersecurity perspective, it requires gathering information from different external sources to track the attack behavior and link it to a single threat actor. 
In this study, we do not consider these objects as our emphasis is on extracting intelligence from individual reports.
Conversely, the SDOs \textit{Malware Analysis}, \textit{Grouping}, \textit{Report}, and \textit{Observed Data} enable a more comprehensive and detailed representation, organization, and reporting of threats, analyses, and observations, offering a broader context beyond just specific malware characteristics.

\textbf{STIX Relationship Objects.} 
\textcolor{black}{To link SDOs, STIX introduces the \textit{Relationship} SRO that has 3 main components: the source object, the target object, and the relationship type. These components are pre-defined by STIX in a relationship matrix that can be extracted from STIX documentation. 
There are 38 relationship types including (but not limited to):
\textit{indicates}, \textit{targets}, \textit{uses}, \textit{exfiltrates to}, \textit{authored by}, \textit{communicates with}, etc. Every pair of SDOs has its own set of possible relationship types. For example, a relationship between a \textit{Malware} object and a \textit{Location} object is either \textit{targets} or \textit{originates from}. It is worth noting that some pairs of SDOs do not have relationships. In addition, STIX defines some relationships between some SCOs and SDOs such as \textit{Infrastructure} and \textit{IPv4 Address}. 
}

\textbf{STIX Cyber-observable Objects.} The \textit{Indicator} SDO 
can have different sub-types which are defined by the SCOs.
These include (but are not limited to)\ \textit{Directory, Domain Name, Email address, File, IPv4/IPv6 address, MAC address, URL}, and  \textit{Windows Registry Key}. We note that an \textit{Indicator} object cannot be defined without defining its sub-type that must be chosen from the SCOs listed in STIX documentation \cite{stix_docs}.




\section{Characteristics of Security Text}
\label{subsec:challenges}

\textbf{Domain-Specific Terminology.} Threat intelligence reports are primarily intended for security experts who develop and maintain a wide range of security services and functions. As such, the text inherently assumes that the reader possesses a sufficient level of domain expertise to comprehend the intricacies of newly discovered threat behavior without providing redundant details that may be accessible elsewhere in the public domain.
For instance, an author of a threat intelligence report may use terms like ``payload," ``malware," and the actual ``malware name" interchangeably in the same section, assuming that the reader understands their interrelation. Similarly, a threat actor may be referred to by various names assigned by different security vendors.
Furthermore, the author may establish implicit links between entities. For example, when discussing a malware that targets IIS or is written in \textsc{C\#}, the assumption is that the reader is familiar with \textsc{.NET} and ASP, as well as the deployment of \textsc{.aspx} files on an IIS server.
These nuances present challenges to the effectiveness of conventional NLP pipelines and language models, which are mainly built on natural language texts intended for a broad audience.
This mismatch becomes evident in the form of a distribution shift, as the characteristics of the data used to build such models significantly differ from those of the data used for testing. Consequently, this disparity leads to inferior performance. Addressing these issues becomes vital to enhance the adaptability and overall performance of the models when dealing with specialized threat intelligence reports.


\textbf{Language Complexity.} The primary competency of the author of a threat intelligence report lies in their ability to collect and analyze evidence crucial for understanding threat behaviors. 
Communicating these findings to other experts does not necessarily demand a concise and crisp writing style.
In fact, as demonstrated by Satvat et al. \cite{satvat2021extractor}, threat intelligence reports tend to be quite verbose and lack a specific grammatical structure.
Furthermore, they often lack proper punctuation and may contain sentences with missing subjects, objects, or pronouns. 
The cumulative effect of these complexities makes understanding the context of events challenging for LLMs
and renders any inferences derived from such text error-prone.

\textbf{Use of Mixed Data Formats.} Security texts are diverse, containing various types of data such as tables, lists, code snippets, command line arguments, figures, and charts, in addition to the main textual content. Authors often incorporate these different types of data directly into sections without clear delimiters. For example, code snippets or commands may be embedded within sections as regular text, lacking code boxes or syntax highlighting.
Tables and charts are employed when information can be conveyed more effectively than using text alone. As a result, to extract comprehensive knowledge from a threat intelligence text, it becomes crucial not only to understand the textual content but also to transform these alternative data formats into text as part of a pre-processing step. 
For instance, some charts show the attack timeline and the employed techniques or the evolution of malware variations. In addition, some figures contain snippets from the reversed binaries which may contain crucial information like malware kill switches. Moreover, command lines and code snippets may contain valuable information that are usually hardcoded in the malware's source code such as private keys, encryption algorithms, domain names, etc. Hence, it is important to consider these artifacts as they convey information that might be helpful in containing malware.





\textbf{Entity Naming Inconsistency.} Entities can exhibit inconsistent naming conventions. For example, an author may use a malware name, its antivirus (AV) detection name, and different variations of the malware name (e.g., uppercase and lowercase) interchangeably. In addition, threat actors might be known with different names that could be all used interchangeably in a single article. Consequently, a single entity may be extracted as multiple separate entities.
To address these ambiguities and ensure accuracy, it is crucial to perform an entity resolution step before undertaking more complex knowledge extraction tasks.

\section{System Overview}
\label{sec:system_overview}
A threat analysis report can be effectively represented as a knowledge graph, where nodes signify threat-related entities and edges represent the relationships between those entities. The goal of our system is to analyze a threat report to extract all mentioned entities, identify their types, determine pairs of related entities, and accurately establish the relationship that connects each pair.
The entity extraction task is typically framed as a named entity recognition task \cite{wang2023gpt,monajatipoor2024llms}. While identifying entity types, and related entities, and extracting relationships may seem like straightforward classification tasks, the complexity lies in accurately interpreting the nuances of security text.
To achieve this objective, we leverage post-trained, general-purpose LLMs and apply continual fine-tuning to adapt them for threat knowledge extraction tasks.

\subsection{System Components}
\begin{figure*}[!ht]
    \centering
    \includegraphics[width=6.8in]{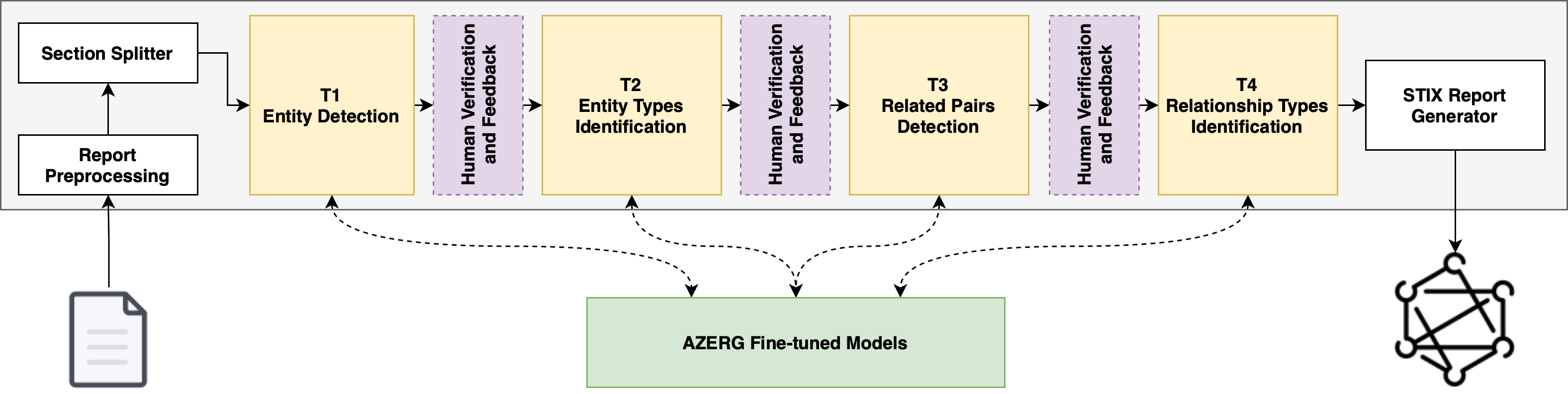}
    \caption{The architecture of \system.}
    \label{fig:azerg_architecture}
\end{figure*}

Figure \ref{fig:azerg_architecture} presents the system architecture alongside a depiction of the data workflow.
The system preprocesses CTI text and generates input segments, which are then sequentially processed by the fine-tuned models. 
The output from T1, the entity detection stage, is first verified and then passed to T2 for entity type classification. Once T2 produces typed entities, they are verified and forwarded to T3, along with the original text segment and STIX relationship matrix constraints. T3 identifies potential related entity pairs, which, after verification, are passed to T4 for relationship type classification using contextual information. Finally, the verified entities and their relationships are compiled by the STIX Report Generator to produce the structured output.

\textbf{Report Preprocessing.} 
This component accepts input in the form of a URL, PDF document, or text document, and converts it into plain text. In case of HTML, it strips away all formatting except for HTML heading tags, which are commonly used for titles.
Given the variety of data found in CTI reports, which includes both textual and non-textual content, preprocessing of these reports is essential before undertaking extraction tasks.
During preprocessing, we retain all code snippets and command lines due to their potential relevance in providing insights into attack entities and threat behaviors, especially given that many modern LLMs can interpret code. Additionally, images, figures, charts, and graphs are crucial non-textual elements frequently found in CTI texts. Extracting information from these visual elements exceeds the capabilities of LLMs, requiring the use of specialized vision models. We leave the extraction of cyber security entities and relationships from visual elements as future work.

\textbf{Section Splitter.} This component processes plain text reports by dividing them into segments. Threat intelligence reports are typically structured into sections, each dedicated to a specific aspect of the content. We utilize the section titles to segment these reports. This partitioning strategy ensures that entities associated with particular threat activities are grouped closely, providing the model with the necessary context to execute each task effectively.

\textbf{Fine-Tuned Models.} Our initial exploration of several well-established instruction-following models, that are fine-tuned across a broad spectrum of tasks spanning numerous capability areas, revealed that their performance varies across the four tasks. 
This indicated that some of our tasks are more complex and require specialized knowledge. 
Consequently, we opted for continual fine-tuning of these models. 
During fine-tuning, we incorporate the task-specific prompt with a context that includes information about STIX entities and relationships pertinent to the task, along with examples
In our strategy, we considered both developing task-specific models and a model specializing in the combined tasks. 
We trained these models using a specially curated dataset necessary for the four tasks (Sec. \ref{subsec:datasets}). 
We assessed the impact of the varying learning rate, temperature, and top-p values to determine the optimal hyperparameter setting (Sec.~\ref{sec:prompting_strategy}).

\textbf{Entity Detection (T1).} This module performs the task of detecting all STIX entities (i.e., SDOs and SCOs) mentioned in the text passage. Identifying all SDOs and SCOs in a report is essential for generating a comprehensive STIX output. When dealing with SDOs, our primary focus lies in identifying the names of the following SDOs: \textit{Attack Pattern, Identity, Location, Malware, Threat Actor, Campaign, Tool, Infrastructure}, and \textit{Vulnerability}, the descriptions of \textit{Course of Action} objects, and the value and sub-type of \textit{Indicator} objects. Regular expressions have proven highly effective for accurately detecting structured attack indicators, such as IP addresses, hashes, YARA rules, and registry keys. For T1, we begin by identifying indicators using Indicators of Compromise extractors, explained in Sec. \ref{subsec:implemention}, before applying our models.


\textbf{Entity Types Identification  (T2).} The purpose of this module is to identify the type of each entity detected by T1, based on the context in which the entity is mentioned. The identification of entity types is crucial for constructing the final STIX graph, as this information determines the potential relationships between entities. This module takes as input the identified entities in T1 and the text passages where they were mentioned. Then, it goes through them individually and asks the model to identify their STIX types.


\textbf{Related Pairs Identification (T3).}
Given a list of entities and their types along with the text passage where they are mentioned, this module identifies pairs of related entities. It utilizes the STIX relationship matrix, which defines valid pairwise relationships (SROs) among all entity types (i.e., SDOs and SCOs). By iterating over all pairs of entity types between which an SRO can be defined, the module extracts all possible entity pairs that can be connected through a valid SRO. For each pair of entities, multiple relationship options are provided to the model to determine valid relationship types that describe their interaction. To assess whether two entities are related, these choices are supplemented with two additional options: \quotes{is not related to} and \quotes{not sure}.
This process is repeated until all SROs associated with valid entity pairs are identified.


\textbf{Relationship Types Identification (T4).} This module takes as input all pairs of entities with identified types, a list of possible relations between each pair, and a text passage where both entities are mentioned, then it determines the most likely relationship between a pair of related entities. 


\textbf{Human Verification and Feedback.} 
Given that the tasks follow a sequential order, a human verification and feedback phase is incorporated following each task. This allows users to add, delete, or alter the output from each task before it is used in the subsequent one.

\begin{figure}[!ht]
    \centering
    \begin{tcolorbox}[
        colback=gray!4, 
        colframe=black,
        boxrule=0.5pt,
        left=5pt,
        right=5pt,
        top=5pt,
        bottom=5pt,
        boxsep=0pt,
        width=\linewidth
    ]
    \fontfamily{ptm}\selectfont\small
        The cyber espionage group known as \quotes{Shadow Dragon}, notorious for targeting financial institutions, recently conducted a campaign against "Global Finance Corp." This institution, primarily located in United States, was targeted using a custom malware variant called \quotes{SerpentStealth}. The initial infection vector leveraged the \quotes{Spearphishing Attachment} technique (MITRE ATT\&CK T1566.001), tricking employees into executing the payload. Once active, SerpentStealth established communication with a command-and-control (C2) server located at the IP address \textsc{198.51.100.5} for data exfiltration and receiving further commands.
    \end{tcolorbox}
    \vspace{\baselineskip}
    \includegraphics[width=3.2in]{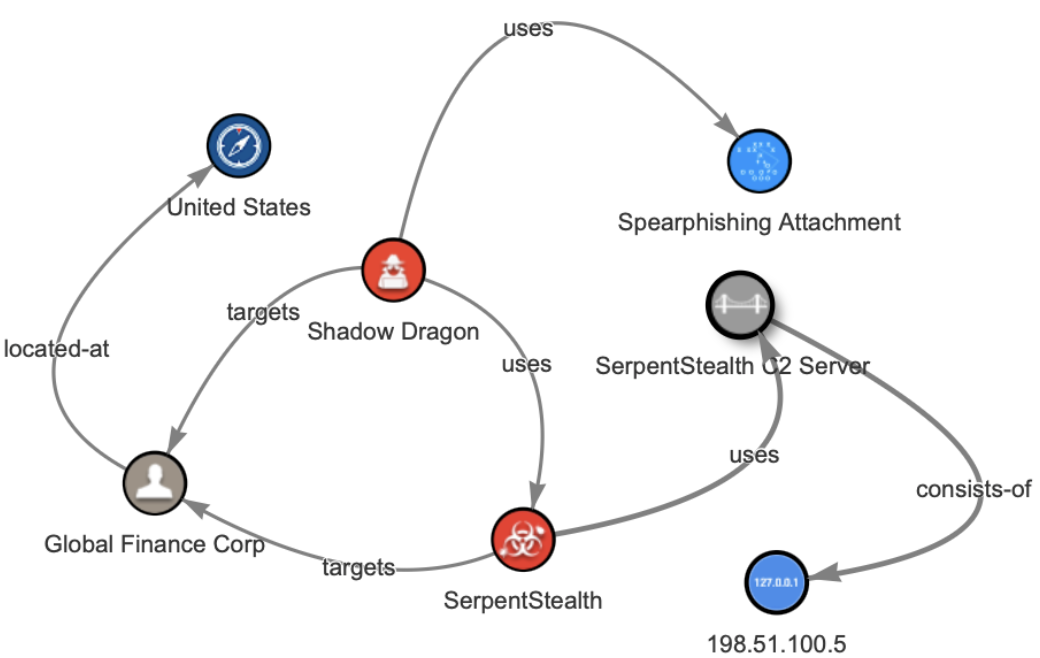}
    \caption{Sample excerpt and its corresponding STIX graph representation.}
    \label{fig:frebiis_graph_example}
\end{figure}

\textbf{STIX Output Generator.} Upon identifying the entities and extracting relationships following the STIX standard, we merge them to generate a JSON file that encompasses entities and their relations in STIX format. This file serves as a means for threat intelligence sharing and can be utilized with TAXII\footnote{\url{https://oasis-open.github.io/cti-documentation/taxii/intro}} or any other threat intelligence exchange protocol. 
Figure~\ref{fig:frebiis_graph_example} illustrates an example of the extracted knowledge graph generated from user-provided text.


\section{Datasets}
\label{subsec:datasets}

Fine-tuning and testing models for threat knowledge extraction require ground truth annotations that align with STIX standard definitions.
Many threat intelligence vendors supply STIX reports to their subscribers alongside threat analysis reports. However, our search for publicly available STIX report data revealed significant limitations. These STIX reports primarily consist of indicator-type SDOs, which are largely detectable through regular expression-based methods, but lack essential SROs that define the relationships between entities. 
To address this need, we manually curated a new dataset of STIX entity and relationship objects.
This dataset comprises two sources, which we refer to as AZERG Data and AnnoCTRPlus—a revised and augmented version of the AnnoCTR dataset introduced in \cite{lange2024annoctr}.

\textbf{AZERG Data.} It comprises SDOs and SROs extracted from 21 {\it full} reports on malware campaigns recently published by 11 threat intelligence vendors.
Our curation of these reports was guided by strict selection criteria, aiming for a thorough evaluation of LLM capabilities across diverse report types. We specifically focused on selecting reports that included a diverse range of textual elements—such as command lines and code snippets—that extend beyond what regular expressions can detect. The average report length is 1650 words, with individual reports varying between 757 to 3.4K words.

\textbf{AnnoCTRPlus.}
A portion of the AnnoCTR dataset annotates named entities—such as organizations, locations, industry sectors, code snippets, hacker groups, malware, tools—as well as time expressions and adversarial tactics and techniques, across 120 real-world CTI reports from five vendors.  Because some of these annotations overlapped with defined entity types in the STIX standard, we manually reviewed and expanded them to ensure alignment with the STIX standard trough a two step process.
\begin{itemize}
\item Entities not belonging to any defined SDO category (referred to as Concepts in AnnotCTR), such as the words \textit{malware}, \textit{attack}, and \textit{payload}, were removed, while missing STIX entities like \textit{indicators}, \textit{vulnerabilities}, and \textit{courses of action} were added. The \textit{indicator} type enttities were added automatically using regular expressions, while others, such as \textit{course of action}, \textit{campaign}, and \textit{infrastructure}, required manual annotation.
In the AnnoCTR dataset, each entity in multi-entity sentences was initially referenced separately. We consolidated these entries using fuzzy string matching, reducing unique text passages to 744.
Additionally, the dataset includes inferred MITRE ATT\&CK tactic and technique IDs, even when not explicitly specified in the text. Due to the challenges of extracting MITRE ATT\&CK techniques from security texts \cite{10.1145/3634737.3645000,nguyen-srndic-neth-ttpm}, we retained only explicitly mentioned \textit{attack patterns}.
\item The AnnoCTR dataset lacks entity relationships, so the SROs within its text collection were manually annotated.
\end{itemize}

The manual annotation of both datasets was performed by an expert in offensive security with over ten years of experience in malware analysis. This expert conducted an in-depth analysis of the STIX standard and thoroughly annotated all SDO and SCO entities, along with their SRO relationships, within the threat intelligence reports using the Doccano framework\footnote{\url{https://doccano.github.io/doccano/}}.
To ensure accuracy, two additional experts—one co-author and an external expert with comprehensive knowledge of the STIX standard—reviewed these annotations and performed cross-verification to resolve any inconsistencies.
Disputes that arose were mainly in three areas: (1) determining the appropriate STIX relationship type for certain malware actions. For instance, deciding if a malware \textit{checking} for a tool's existence in a system aligns with the \textit{uses} relationship due to lack of more fitting relationships; (2) differentiating between similar relationship types, such as \textit{owns} versus \textit{hosts}, \textit{downloads} versus \textit{drops}, and \textit{uses} versus \textit{exploits}; and (3) clarifying entity types in cases of ambiguous naming, particularly concerning \textit{malware} and \textit{threat actor} entities. We provide the details of our two sources in Table \ref{tab:data_sources_information}.

\begin{table}[!ht]
\centering
\caption{Details of our data sources.}
\scalebox{0.75}{
\begin{tabular}{lcccc}
\toprule
\textbf{Dataset} & \textbf{Text Passages} & \textbf{Avg. Word per Passage} & \textbf{Entities} & \textbf{Relations} \\ \hline
AZERG Data       & 170                    & 102.76                              & 2041              & 1073                \\ 
AnnoCTRPlus      & 744                    & 23.56                               & 1970              & 1002               \\ 
\bottomrule
\end{tabular}
}
\label{tab:data_sources_information}
\end{table}

\textbf{Train and Test Splits.} 
The dataset is divided into two non-overlapping parts for training and testing at report and campaign levels. Although AnnoCTRPlus includes 120 CTI reports compared to AZERG’s 21, the resulting number of entities and relationships was comparable, suggesting AnnoCTR used shorter reports and partially annotated the reports, potentially capturing fewer complex objects and relationship types. Additionally, the text passages containing these STIX objects were noticeably shorter (see third column of Table \ref{tab:our_dataset_information}), limiting contextual detail. To accurately assess the real-world performance of our tool in terms of precision and recall for entities and relationships within \textit{full} reports, we designated 11 AZERG Data reports for testing and 10 for training. This split also allowed for vendor-level separation: the training set included reports only from three vendors, while the test set comprised reports from eight non-overlapping vendors. To prevent any potential contamination, we further ensured that malware campaigns described in the training set were absent from the testing set.

Overall, the training split contains 2,664 entities and 1,510 entity relationships across 806 text passages (i.e., sections) that provide context for detecting these objects within 130 CTI reports.
The test split includes 108 text passages, containing 1347 entities and 565 relationships, representing 33.58\% of all the entities and 27.22\% of all relationship objects. To the best of our knowledge, this is the largest publicly available dataset annotating entities and relationships according to the STIX standard\footnote{The dataset and models can be found at: \url{https://huggingface.co/collections/QCRI/azerg-687264a76236a362e833d8eb}}.

\begin{figure}[!ht]
    \centering
    \scalebox{0.28}{
    \includegraphics{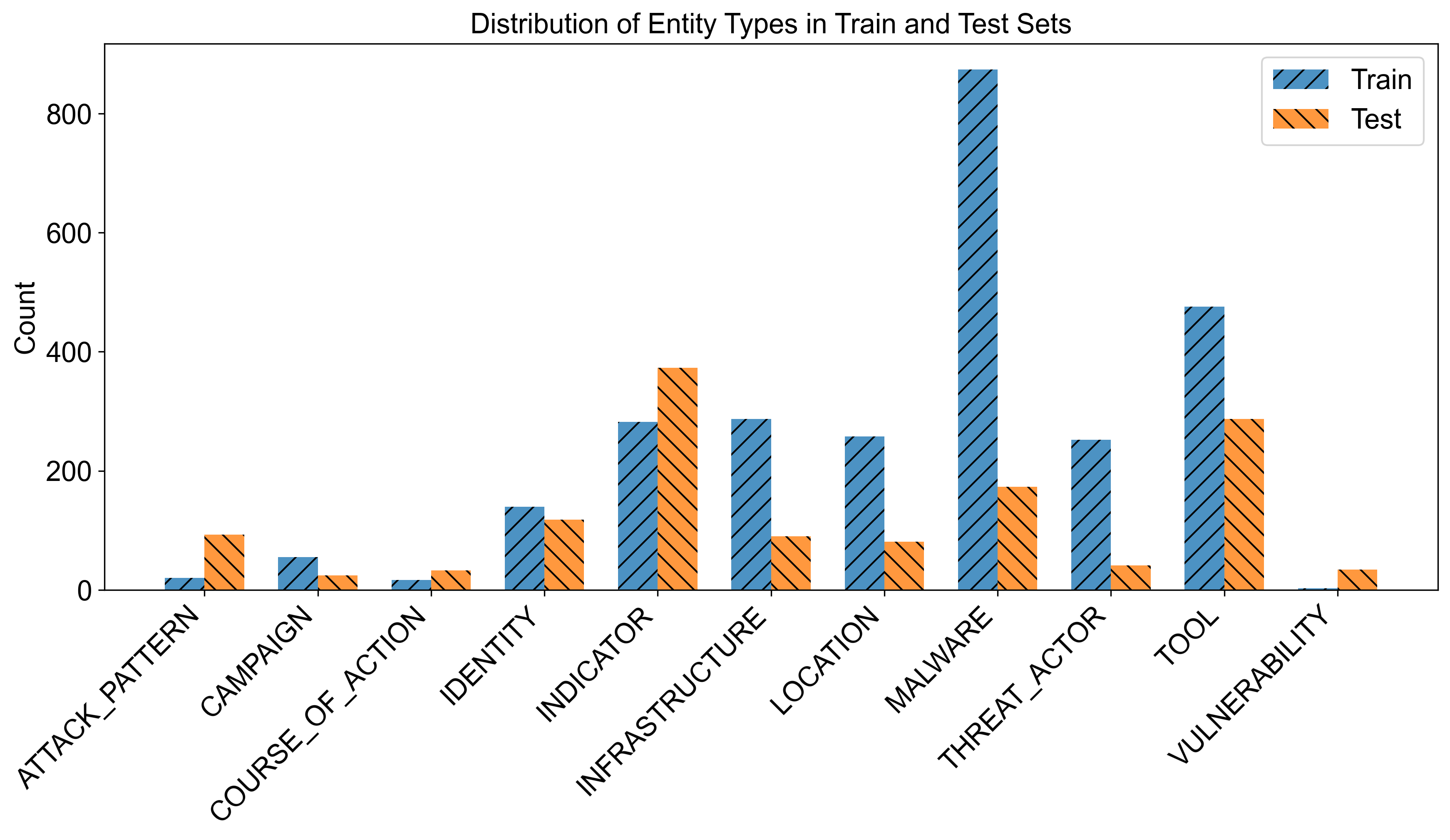}
    }
    \caption{Entity type distribution in our dataset.}
    \label{fig:entities_distribution}
\end{figure}

\begin{figure}[!ht]
    \centering
    \scalebox{0.28}{
    \includegraphics{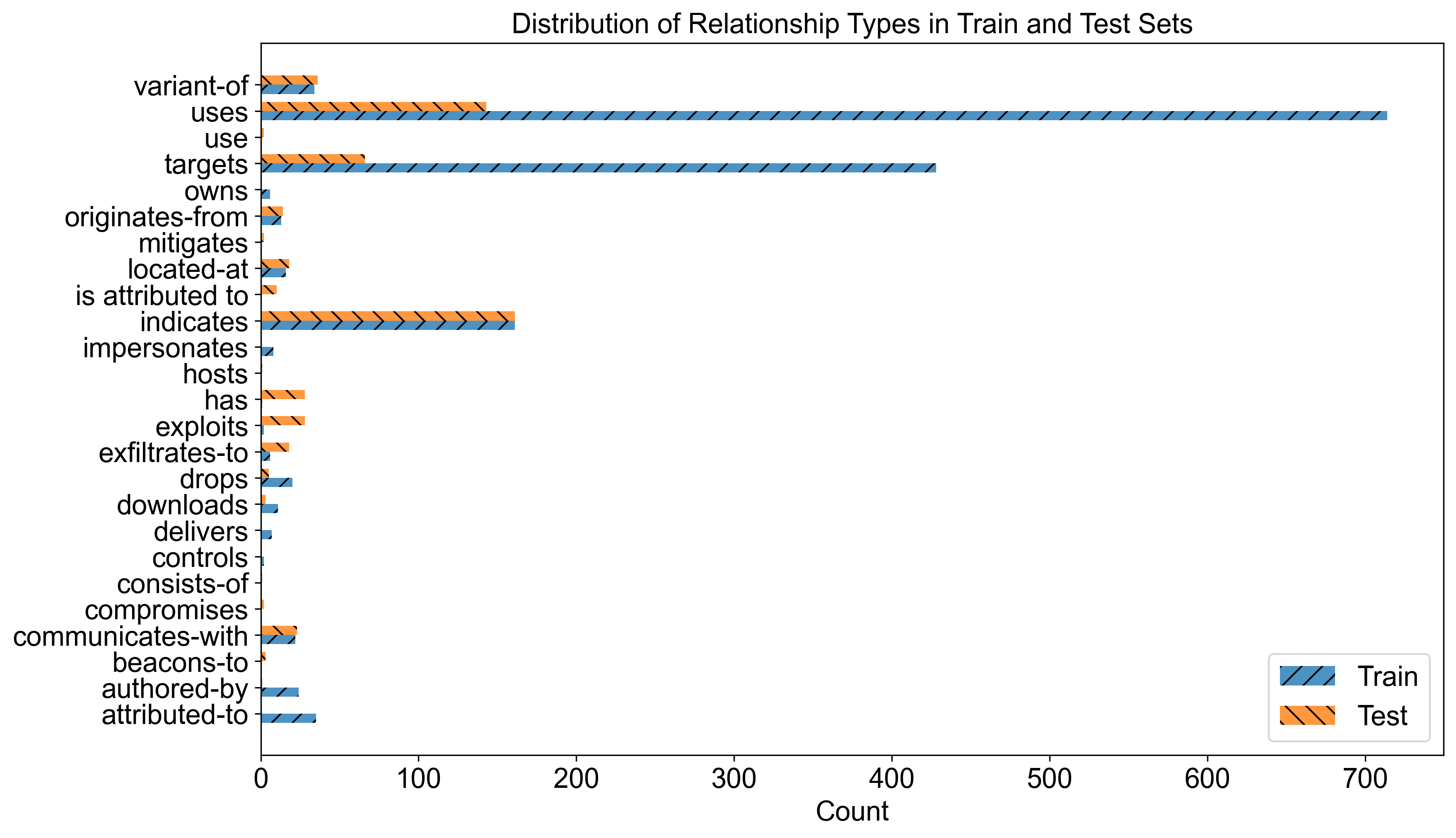}
    }
    \caption{Relationship type distribution in our dataset.}
    \label{fig:relations_distribution}
\end{figure}

Table \ref{tab:our_dataset_information} provides the total number of STIX annotations generated, along with the text contexts containing these STIX objects.
Figures \ref{fig:entities_distribution} and \ref{fig:relations_distribution} display the distribution of entity types and relationships in our Train and Test splits, respectively.

\begin{table}[!ht]
\centering
\caption{Details of our train and test splits.}
\scalebox{0.77}{
\begin{tabular}{lcccc}
\toprule
\textbf{Dataset} & \textbf{Text Passages} & \textbf{Avg. Word per Passage} & \textbf{Entities} & \textbf{Relations} \\ \hline
Train      & 806                    & 35.5                              & 2664              & 1510                \\ \hline
Test       & 108                    & 244.0                               & 1347              & 565               \\ \hline
Total            & 914                    & 59.69                               & 4011              & 2075               \\
\bottomrule
\end{tabular}
}
\label{tab:our_dataset_information}
\end{table}

\section{Model Fine-Tuning}
\label{sec:prompting_strategy}

LLMs typically undergo post-training with custom datasets to enhance their ability to respond to a diverse array of instructions. These post-trained models, often referred to as chat or instruct models, become highly skilled at understanding human intentions and executing required tasks. However, these models may underperform in more specialized tasks where the pre-training corpus lacks sufficient task-related data or when the tasks deviate significantly from those encountered during the post-training phase. In such cases, rapid adaptation to a specific task can be achieved by incorporating a small number of examples directly into the task instruction, a method known as few-shot prompting. In our evaluation of various open- and closed-parameter models, including GPT4o and Mistral-7B-Instruct-v0.3, we observed that they can execute tasks T1-T4 with F1-scores ranging from 0.15 to 0.77 (Further details on these results are provided in Sec. \ref{sec:evaluation}).


In our approach to fine-tuning, we performed continual fine-tuning on already post-trained, open models to leverage their existing capabilities.
To achieve this, we used our curated dataset of STIX annotations described in Sec. \ref{subsec:datasets}.
Utilizing this data, we developed both task-specific models for T1-T4 individually, and a more comprehensive model that specializes in all four tasks.

To select the final base model for task adaptation, we fine-tuned six instruction-tuned models using AZERG Data: Google’s Gemma-2-9b-it, Alibaba’s Qwen2-7B-Instruct, Meta’s Llama-3.1-8B-Instruct, Shanghai AI Lab’s InternLM2\_5-7b-chat, Mistral AI’s Mistral-7B-Instruct-v0.3, and Microsoft’s Phi-3-mini-instruct.
These models exhibited varying performance on our tasks, with Gemma-2-9b-it initially performing the best. However, after further fine-tuning using our task data, the Mistral-based model outperformed the others, leading us to select it as our final base model.

The fine-tuning strategy for each task-specific model varied in terms of how the models were prompted and the context provided. 
For tasks T1 and T2, we supplied the model with definitions of all STIX entity types and examples of entities where applicable. The prompts for these tasks were designed to detect the names of entities (T1) and to identify the types of entities extracted during T1 (T2). Similarly, for tasks T3 and T4, we included possible entity relationship types between each pair while adding the \quotes{is not related to} and \quotes{note sure} options. In each instance, the model was tasked with determining whether a pair of entities are related (T3) and identifying the type of relationship between them (T4). The prompt templates for each task are presented in Appendix \ref{apdx:prompt_templates}. 



For fine-tuning, we used the training split of our curated dataset (Table \ref{tab:our_dataset_information}), following approximately a 70/30 split ratio. Notably, the training and testing reports were completely non-overlapping. Additionally, the vendors of the reports and the campaigns featured in the testing set were excluded from the training data to prevent data contamination and to more accurately evaluate the real-world performance of our system.

\section{Evaluation}
\label{sec:evaluation}

\subsection{Implementation}
\label{subsec:implemention}

For report preprocessing, we developed a custom parser that segments reports into sections using heading tags, leveraging BeautifulSoup\footnote{\url{https://www.crummy.com/software/BeautifulSoup/}}. In instances where heading tags are absent or sections are excessively long, we segment the text based on a predetermined length, allowing for overlapping segments. This approach was generally unnecessary with high-quality vendor reports but was implemented to accommodate PDF reports, which we parse using PyPDF\footnote{\url{https://github.com/py-pdf/pypdf}}. For matching \textit{Indicator} objects in text, we utilized IoCFinder\footnote{https://pypi.org/project/ioc-finder/} and IoCParser\footnote{https://pypi.org/project/iocparser/} libraries.


For fine-tuning, we employed LLaMa-Factory\footnote{\url{https://github.com/hiyouga/LLaMA-Factory}} with Low-Rank Adaptation (LoRA) \cite{hu2021lora}, and a learning rate set at $10^{-4}$. Regarding sampling parameters, we use a \textit{temperature} of 0.7, \textit{top-p} of 0.1, and a maximum number of tokens is 1024 for T1, 10 for T2, T3, and T4. These parameters were used for all the experiments and all the models unless otherwise specified.
In Appendix \ref{subsec:parameters_effect}, we detail our selection of parameters following an extensive search. 
The fine-tuning process was conducted on a machine equipped with an NVIDIA A100-80GB GPU, running Ubuntu 22.04.
For inference, we deployed the open models on vLLM\footnote{\url{https://github.com/vllm-project/vllm}}, on one NVIDIA V100 32GB GPU. To interface with the models and guide their outputs, we employed DSPy\footnote{\url{https://dspy-docs.vercel.app}}, where prompts are encoded as \textit{Signatures}—DSPy's term for declarative specifications that outline the expected input and output, thereby helping the models comprehend the tasks they need to perform.


\subsection{Performance of Fine-Tuned Models}
\label{subsec:entities_extraction_eval}

We assess the performance of our fine-tuned models in tasks T1-T4 (Sec. \ref{sec:system_overview}). During testing, each text passage from the test split is used individually for extraction. The models are prompted to perform each task using the same template that was employed during fine-tuning. The responses from the models are compared to the ground truth data to calculate precision, recall, and F1-score values. We report results for models specifically fine-tuned for each task (referred to as \textit{AZERG-S-T*}, where the star represents the task number), as well as for our AZERG-MixTask model, which was fine-tuned using a combination of the same data used for the specialized models. Additionally, we present comparative results from post-trained LLMs such as GPT4o and Mistral, alongside state-of-the-art methods for each task. 

Table \ref{tab:results_T1} provides the performance for entity detection (T1), a common task in threat knowledge extraction. For this task, we evaluated two state-of-the-art methods, \cite{attackg:esorics:2022, satvat2021extractor}, and also included a generic transformer-based model for named entity recognition \cite{zaratiana2023gliner} to benchmark our performance. We note that the Regular Expressions were integrated with our models and all the approaches that we compare with to ensure a fair comparison. 

Similarly, Table \ref{tab:results_T2} presents the results for the entity-type detection task (T2). Since T2 is not addressed by other methods, we only include results achieved by LLMs. In both cases, we observe that our fine-tuned models significantly outperform the other approaches, by a margin of 20\% or more. Notably, for T1, the mixed-task model yielded considerably better results (84.43\%) compared to the task-specific model (80.23\%). For T2, the task-specific model performed marginally better (89.23\% vs 88.49\%).

The performance metrics for entity relationships are presented in Tables \ref{tab:results_T3} and \ref{tab:results_T4} for tasks T3 and T4, respectively.
For T3, our fine-tuned models outperformed generic instruction models, with the mixed-task model achieving a 2.3\% improvement over the best-performing GPT4o model. 
Additionally, the mixed-task model surpassed the task-specific model by 1.5\% (95.47\% vs. 93.97\%).
For T4, our fine-tuned models demonstrated significantly better performance than the best-performing GPT4o model, with a margin of 13-14\%. The mixed-task model also outperformed the task-specific model by 1.3\%.
Deploying the AZERG-MixTask model in combination with AZERG-S-T3 is expected to deliver optimal system performance. However, given the marginal difference in T3 performance, the mixed-task model offers a satisfactory alternative with significantly lower computational cost.

Overall, the results demonstrate that our fine-tuned models achieve an F1 score of approximately 84\% or higher across all threat knowledge extraction tasks, with T1 and T4 presenting greater challenges.
The difficulty of T1 likely stems from its open-ended nature, requiring the model to identify all entities within a passage—a task that becomes increasingly complex as the number of entities grows. For T4, the challenge appears to arise from the need to classify among similar relationships (see Sec. \ref{subsec:error_analysis}).
Additionally, the results indicate that only on T3 does the GPT4o model perform comparably to our fine-tuned models, while a significant performance gap is observed for other tasks.


\begin{table}[!ht]
\centering
\caption{Evaluation of Model Accuracy and Comparison for T1}
\scalebox{1.0}{
\begin{tabular}{llll}
\toprule
\textbf{Approach}   & \textbf{Precision} & \textbf{Recall} & \textbf{F1-Score} \\ \midrule
GPT4o           & 0.8635 & 0.4930 & 0.6277 \\ 
Mistral-7B-Instruct-v0.3       & 0.7104 & 0.5003 & 0.5871 \\ 
AttaKG \cite{attackg:esorics:2022}          & 0.3797      & 0.4098    & 0.3941       \\ 
EXTRACTOR \cite{satvat2021extractor}        & 0.2640      & 0.3537   & 0.3023      \\ 
GliNER \cite{zaratiana2023gliner}          & 0.2315      & 0.1997    & 0.2159     \\ \midrule
AZERG-S-T1   & 0.8482 &  0.7611 & 0.8023  \\  
AZERG-MixTask             & \textbf{0.9092} & \textbf{0.7880} & \textbf{0.8443} \\ \bottomrule
\end{tabular}
\label{tab:results_T1}
}
\end{table}

\begin{table}[!ht]
\centering
\caption{Evaluation of Model Accuracy and Comparison for T2}
\scalebox{1.0}{
\begin{tabular}{llll}
\toprule
\textbf{Approach}   & \textbf{Precision} & \textbf{Recall} & \textbf{F1-Score} \\  \midrule
GPT4o        & 0.6481            & 0.6481         & 0.6481           \\ 
Mistral-7B-Instruct-v0.3      & 0.3363             & 0.3363         & 0.3363           \\ \midrule 
AZERG-S-T2                     & \textbf{0.8923} & \textbf{0.8923} & \textbf{0.8923} \\  
AZERG-MixTask             & 0.8849            & 0.8849          & 0.8849            \\ \bottomrule
\end{tabular}
\label{tab:results_T2}
}
\end{table}

\textbf{Inference Time.} For tasks that require more complex generation, such as extracting all entities from a given text (T1), the inference time is naturally higher at 2.57 seconds per query. This reflects the increased computational demands of this more involved task.
In contrast, for tasks with more streamlined output requirements, such as the other three tasks (T2, T3, and T4), the inference times are significantly faster, ranging from 1.54 seconds per query for T2 down to 0.58 and 0.36 seconds per query, for T3 and T4, respectively. This demonstrates our system's ability to scale its performance to match the needs of these specific tasks.


\begin{table}[!ht]
\centering
\caption{Evaluation of Model Accuracy and Comparison for T3}
\scalebox{1.0}{
\begin{tabular}{llll}
\toprule
\textbf{Approach}   & \textbf{Precision} & \textbf{Recall} & \textbf{F1-Score} \\ \midrule
GPT4o           & 0.9234             & 0.9398         & 0.9315            \\ 
Mistral-7B-Instruct-v0.3       & 0.8873            & 0.9203         & 0.9035            \\ 
EXTRACTOR  \cite{satvat2021extractor}       & 0.0889            & 0.0917         & 0.0902           \\ 
GliREL  \cite{zaratiana2023gliner}          & 0.7168            & 0.1849         & 0.2939           \\ \midrule
AZERG-S-T3       & \textbf{0.9335}            & \textbf{0.9451}         & 0.9393    \\ 
AZERG-MixTask             & 0.9224            & 0.9893         & \textbf{0.9547}          \\ \bottomrule
\end{tabular}
\label{tab:results_T3}
}
\end{table}

\begin{table}[!ht]
\centering
\caption{Evaluation of Model Accuracy and Comparison for T4}
\scalebox{1.0}{
\begin{tabular}{llll}
\toprule
\textbf{Approach}   & \textbf{Precision} & \textbf{Recall} & \textbf{F1-Score} \\ \midrule
GPT4o            & 0.7946            & 0.7946        & 0.7946           \\ 
Mistral-7B-Instruct-v0.3       & 0.7097            & 0.7097           & 0.7097             \\  \midrule
AZERG-S-T4                  & 0.8335  & 0.8335  & 0.8335  \\  
AZERG-MixTask             & \textbf{0.8460 }           & \textbf{0.8460}          & \textbf{0.8460}            \\ \bottomrule
\end{tabular}
\label{tab:results_T4}

}
\end{table}

\subsection{Error Analysis}
\label{subsec:error_analysis}

To identify the shortcomings of the AZERG-MixTask model, we conducted an error analysis.

\subsubsection{Entity Detection (T1)}

Despite the strong overall performance in T1, the analysis reveals specific areas where nuanced challenges remain, primarily impacting recall (0.7880, corresponding to 291 missed entities). One such area involves fine-grained \textit{semantic ambiguity and contextual interpretation}. 
While the model correctly identified most entities, it occasionally struggled to distinguish STIX entities from closely related technical terms (e.g., algorithm names like \textit{AES}) or descriptive concepts (e.g., \textit{Malware-as-a-Service}), resulting in 108 false positives, which accounted for less than 10\% of the total extractions.
The model’s contextual understanding was generally effective but occasionally led to misinterpretations, such as extracting nationalities like \textit{Iranian} or \textit{Chinese}
as standalone entities, or misclassifying code elements like function names such as \textit{MalwareMain}.

A more significant limitation lies in the model’s F1 score (0.8443), which is primarily constrained by its lower recall (0.7880) compared to its high precision (0.9092). This discrepancy results from 291 entities that the model failed to capture from the ground-truth human annotations.
Areas of reduced recall were most evident in non-IoC categories such as \textit{Tool} names, \textit{Malware} variants, \textit{Threat Actor} aliases, and \textit{Identity} references (e.g., vendors like ESET, organizations such as the FBI, or platforms like GitHub). 
Notably, over 25\% of \textit{Identity} mentions were missed, often due to contextual ambiguity regarding their involvement in the attack.
Several factors contributed to these omissions:
(i) Some entities were embedded within complex narrative structures or referenced indirectly—such as through procedural descriptions or comparisons—making them more difficult to isolate than clearly stated facts;
(ii) The presence of dense, non-linguistic elements like IoCs (e.g., file hashes, IP addresses) interspersed throughout the text may have disrupted the model’s language understanding or diverted attention from key entities; and
(iii) The frequent use of aliases and synonyms in cybersecurity reporting, such as multiple names for threat groups like APT34 or OilRig, posed additional challenges for comprehensive extraction when compared to the reference annotations.

\subsubsection{Entity Types Identification (T2)}
\begin{figure}[!ht]
    \centering
    \includegraphics[width=3.4in]{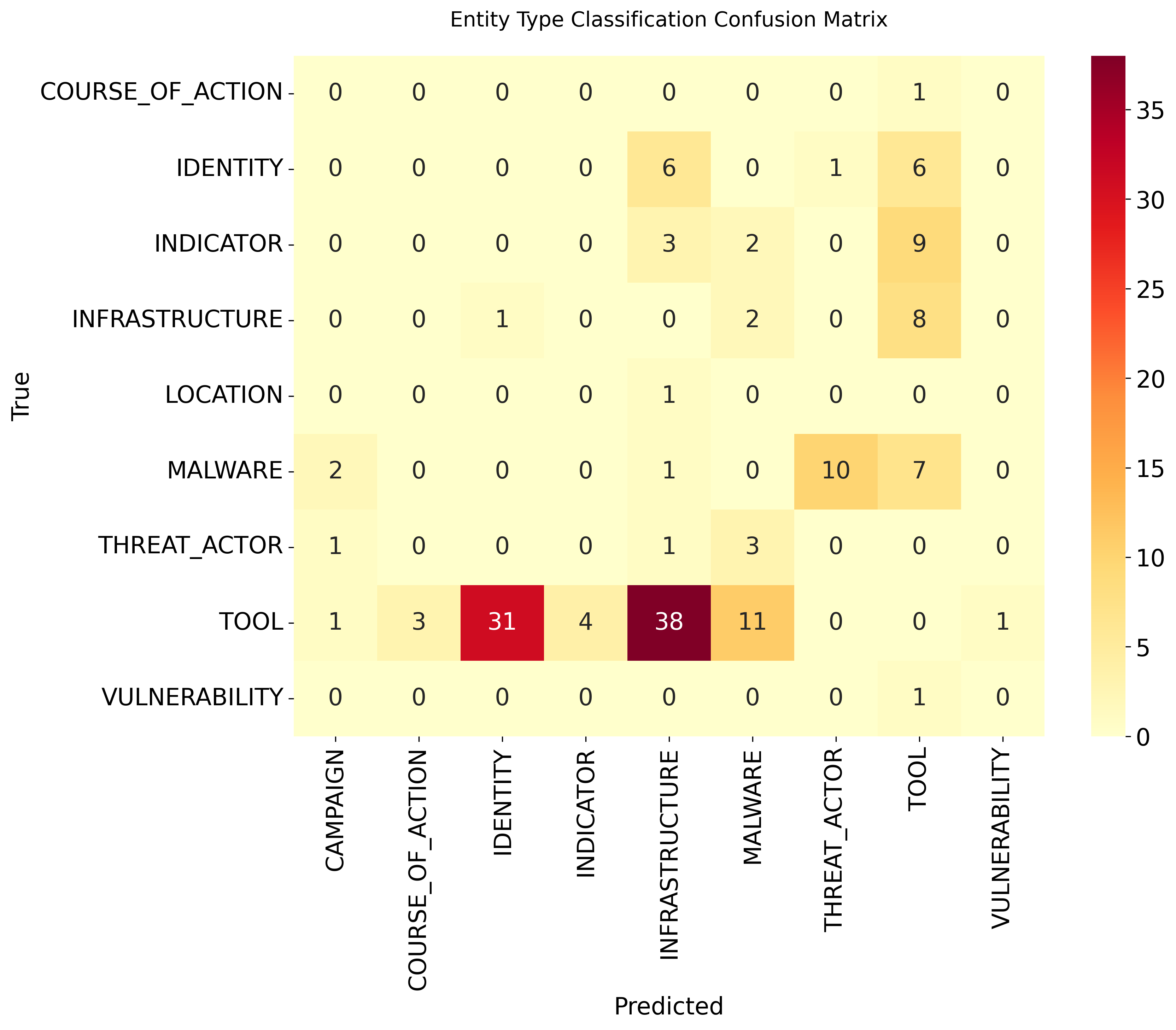}
    \caption{Confusion matrix showing misclassified relationship types in T2.} 
    \label{fig:t2_missed_types_confusion}
\end{figure}

The model frequently mislabeled certain entity types, as shown in Fig. \ref{fig:t2_missed_types_confusion}, often confusing \textit{Tools} with \textit{Infrastructure}, \textit{Tools} with \textit{Identities}, \textit{Tools} with \textit{Malware}, and \textit{Threat Actors} with \textit{Identities}.
For example, distinguishing between \textit{Malware} and a \textit{Threat Actor} often requires additional contextual information that may not always be present in the report. 
In some cases, the \textit{Malware} and \textit{Threat Actor} share the same name, with their classification depending on the context, creating ambiguity for both the models and human analysts.
A similar issue happens when text passages describe VPN servers or virtual machines infrastructure. The models tend to classify such named entities as \textit{Tools} (e.g., VMWare vSphere), even when the context indicates they refer to infrastructure targeted by malware. Difficulties with multi-word names can arise because the tokenizer might split common technical terms, or the model's attention mechanism may fail to span the full entity phrase within complex sentences. Crucially, context is essential not only to correctly detect the entity itself (piecing together split terms or understanding the full span) but also to determine how that entity should be subsequently handled. For example, consider the Tool/Infrastructure confusion: product names like 'VMWare vSphere' can represent both the installable software (Tool) and the resulting managed environment (Infrastructure). Disambiguating this requires nuanced contextual understanding, which the model sometimes misses, especially in shorter text segments. The surrounding text provides the necessary context to decide if the entity should be treated purely as an identity (the name/brand), purely as a tool (the function/software), or as a combination of both identity and tool.

\subsubsection{Related Pairs Detection (T3)}
The model performed well, with only minor errors. The results indicate that 6 relationships were missed, while 47 were incorrectly identified. These incorrect relationships often occur in text passages where verbs link two entities, even though such verbs are not considered valid relationships in STIX. For example, verbs like \quotes{checks}, \quotes{spawns}, and \quotes{steals} were sometimes misinterpreted by the model as relationships.  Furthermore, when paragraphs contained multiple entities—such as \textit{malware}, \textit{threat actors}, and \textit{tools}—the model occasionally failed to capture all relationships between them, leading to incomplete relationship coverage.

\subsubsection{Relationship Types Identification (T4)}
\begin{figure}[!ht]
    \centering
    \includegraphics[width=3.4in]{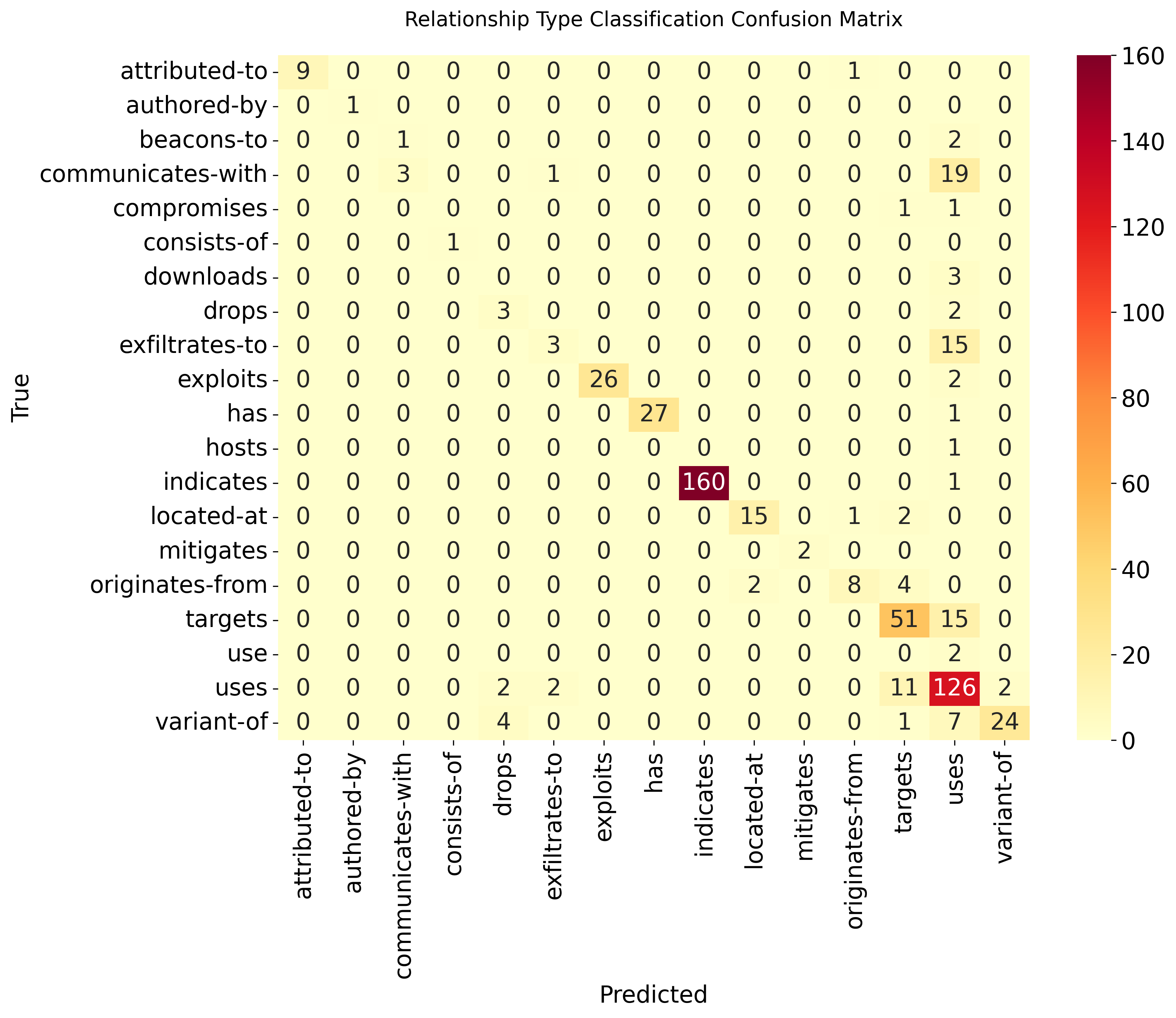}
    \caption{Confusion matrix for Relationship Type Identification (T4) using the AZERG-MixTask model.
    Rows represent the true relationship types, columns represent the predicted types. Darker cells indicate higher counts. Note the confusion between \quotes{uses} (predicted) and other relationships like \quotes{communicates-with}, \quotes{targets}, and \quotes{exploits} (true), highlighting challenges in distinguishing related actions based on text.}
    \label{fig:t4_misclassified_relations}
\end{figure}

The model frequently confuses semantically related relationships, as shown in Fig. \ref{fig:t4_misclassified_relations}. Our analysis reveals that the relationship \quotes{uses} is often misclassified as \quotes{communicates-with}, \quotes{exfiltrates-to}, or \quotes{targets}. While \quotes{uses} is a generic verb that could encompass other actions, the STIX standard provides clear definitions that distinguish it from relationships like \quotes{communicates-with}. 
This discrepancy highlights the challenge of adhering to STIX-specific distinctions. The confusion between semantically similar relationships like \quotes{uses} and \quotes{communicates-with} highlights a challenge in aligning natural language ambiguity with the strict definitions of the STIX standard. Fine-tuning helps, but inherent overlaps in how actions are described textually remain difficult.

\begin{figure}[!htpb]
    \includegraphics[width=3.2in]{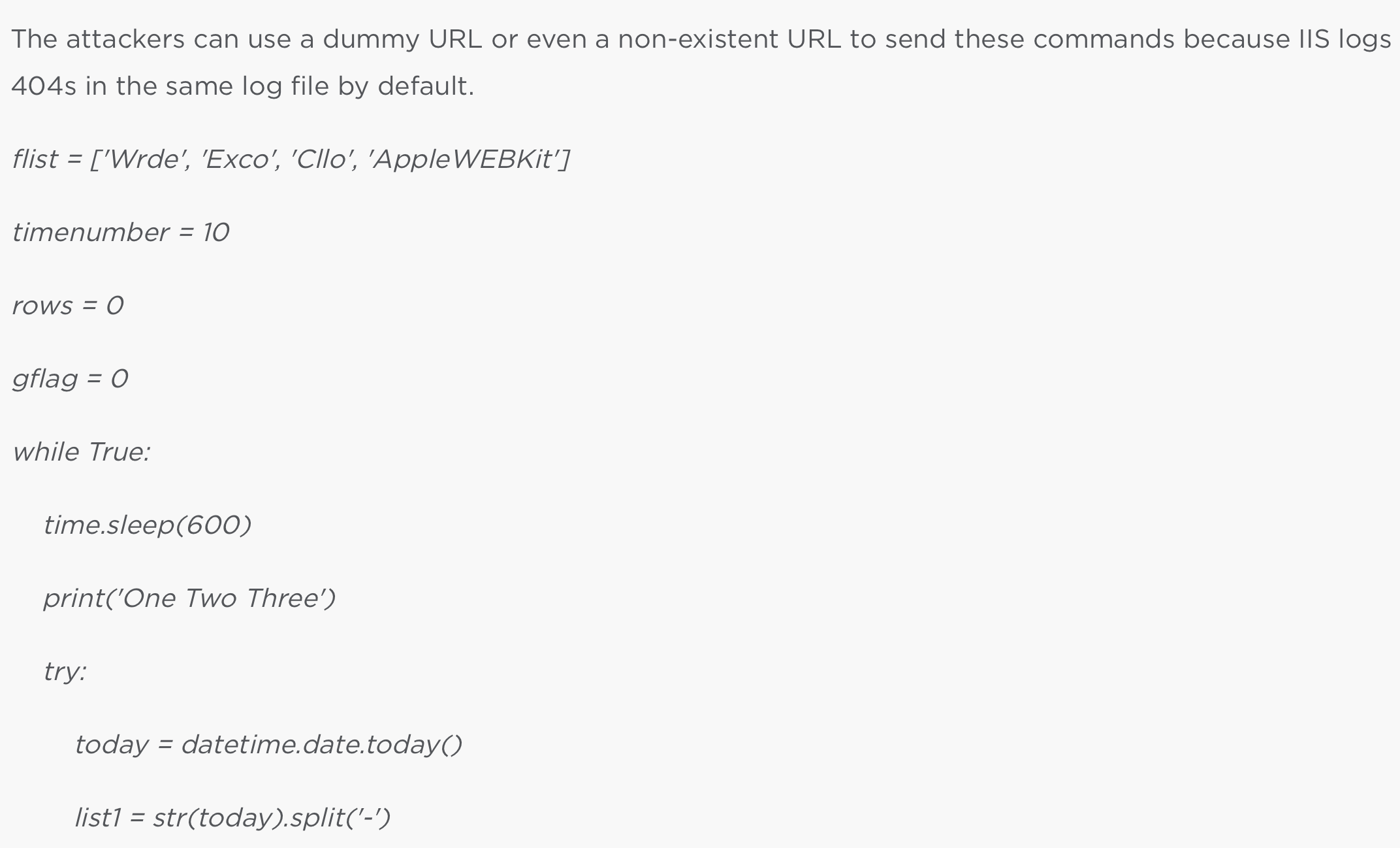}
    \caption{
        \textcolor{black}{
        Example of an error from the Cranefly report \cite{securityCraneflyThreat} encountered during tasks T1 and T2.
        }
    } 
    \label{fig:t1_error_examples}
\end{figure}

\textbf{Error Amplification.}
As shown in Fig. \ref{fig:t1_error_examples}, the presence of Python code within the report caused the model to incorrectly classify the strings \quotes{Wrde}, \quotes{Exco}, \quotes{Cllo}, and \quotes{AppleWEBKit} as \textit{tools}. 
While this error originates from a single misclassification, it gets amplified to multiple strings, significantly degrading the model’s performance in both T1 and T2.

\begin{figure}[!htpb]
    \includegraphics[width=3.2in]{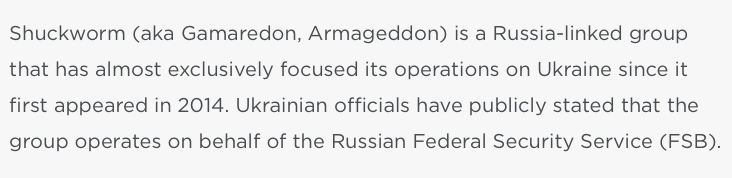}
    \caption{
        \textcolor{black}{
        Example of an error from the Shuckworm report \cite{securityShuckwormInside} encountered during tasks T3 and T4.
        }
    } 
    \label{fig:t3_error_examples}
\end{figure}

In Fig. \ref{fig:t3_error_examples}, the threat actor \quotes{Shuckworm} has aliases \quotes{Gamaredon} and \quotes{Armageddon}. While \quotes{Shuckworm} is correctly identified as having relationships with Russia, Ukraine, and the Russian Federal Security Services, the model fails to detect the same relationships for its aliases, \quotes{Gamaredon} and \quotes{Armageddon}. As a result, the paragraph in the figure alone accounts for 6 missing relationships.

\begin{figure}[!htpb]
    \includegraphics[width=3.2in]{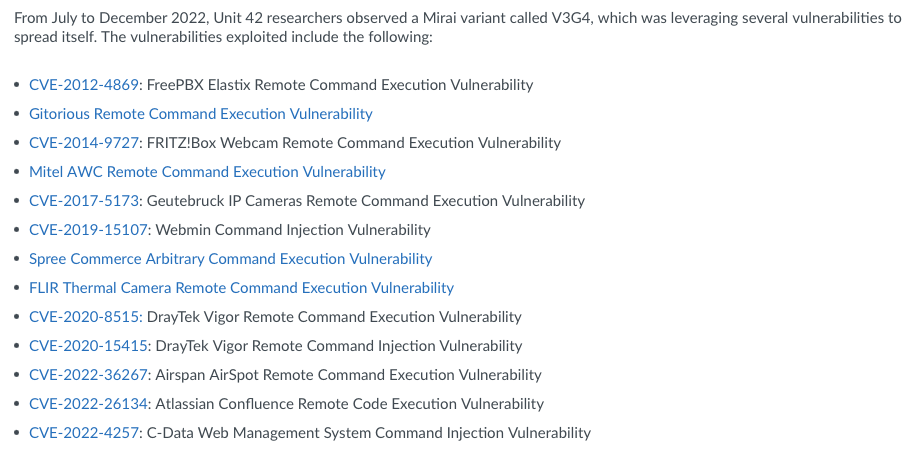}
    \caption{
        \textcolor{black}{
        Example of an error from the V3G4 report \cite{paloaltonetworksMiraiVariant} encountered during tasks T3 and T4.
        }
    } 
    \label{fig:t3_error_examples_2}
\end{figure}

In Fig. \ref{fig:t3_error_examples_2}, the section discusses V3G4, a variant of Mirai, and states that V3G4 exploits 13 vulnerabilities. While the model correctly extracted the relationship between V3G4 and these vulnerabilities, it incorrectly established a relationships between Mirai and the same vulnerabilities, which is not accurate according to the section. The section specifies that V3G4 exploits these vulnerabilities, not Mirai. Additionally, the model erroneously created relationships between tools and vulnerabilities by assigning a single vulnerability to multiple tools.
These examples highlight where our model fails, with the main issue being error amplification as the number of entities in a section increases, which significantly impacts our model's performance, particularly in T3 and T4. However, this kind of mistake correlation also makes it easier for domain experts to identify and correct errors.

\section{Related Work}
\label{sec:related_work}

\textbf{Entity Detection:} This is the most well-studied threat knowledge extraction task. Given that most entities involved in a security context possess well-defined formats, like IoCs, several tools that utilize regular expression rules have been designed for their automated identification.
For entities without standardized formats, like malware family names, adversary groups, and identities, neural-named entity recognition models have been introduced \cite{gao2021system, deka2024attacker, alam2023looking}. These approaches rely on encoder models and finetune them with security text for the Named Entity Recognition (NER) task. More critically, they define their own knowledge graphs rather than the standardized ones like STIX.

\textbf{Entity Relationship Extraction:} Several methods have been proposed with a focus on devising NLP pipelines to identify these relationships accurately.  Most notably, in \cite{gao2021enabling}, a specialized NLP pipeline is used to construct a threat behavior graph. To accurately extract IoC relationships, a dependency parsing-based method is proposed. 
Later, \cite{gao2021system} expanded on this method to identify a wider range of relationships. For this, they modeled the relation extraction task as a multi-class classification problem based on the premise that two entities have a relation when they co-occur within a certain distance within the text. 
To learn these relationships from the text, they employed a piecewise convolutional neural network model with an attention mechanism as the classifier. 

In \cite{satvat2021extractor}, authors introduced EXTRACTOR, a method for extracting a provenance graph from APT attack reports.
A provenance graph represents system entities, such as processes, files, network sockets, etc, as nodes and the operating system calls showing how these entities interact as typed edges. To perform this task, it incorporates a conventional NLP pipeline, involving normalization, resolution, and summarization steps, with a semantic role labeling step to determine the semantic role of sentence components to extract the attack behavior and subject, object, and actions of sentences.

In \cite{li2022attackg}, authors propose AttacKG to aggregate threat intelligence across numerous CTI reports with a focus on attack techniques. This approach involves constructing a knowledge graph that encapsulates the attack workflow at the technique level, as detailed within the CTI reports. Based on this knowledge graph, the authors introduce the \quotes{Technique Knowledge Graph} (TKG), which outlines causal techniques derived from attack graphs, providing a comprehensive depiction of the entire attack chain in CTI reports. The process begins by employing a parsing pipeline to analyze CTI reports, extracting entities relevant to attacks and the dependencies between them to formulate an attack graph. Following this, technique templates are initialized using attack graphs that are built upon examples of technique procedures gathered from the ATT\&CK knowledge base. After that, an improved graph alignment algorithm is employed to correlate technique templates within the attack graphs. This facilitates the alignment and refinement of entities present in both CTI reports and technique templates.

The authors of \cite{gao2022threatkg} propose ThreatKG, which automatically gathers CTI reports from different sources, extracting threat insights, building a comprehensive threat knowledge graph, and enhancing this graph by ingesting new information. To tackle various challenges, ThreatKG proposes a structured hierarchical framework to model a range of entities and relationships in threat knowledge. In addition, it suggests a deep learning-based method for extracting threat intelligence. Moreover, it offers a flexible and expandable system architecture for constructing, maintaining, updating, and exploring the threat knowledge graph.

In \cite{wang2024knowcti}, authors propose KnowCTI which is a tool for extracting cyber threat intelligence. It incorporates cybersecurity knowledge to enhance semantic understanding in the security domain. The process involves building a knowledge base, training knowledge embeddings, refining relevant knowledge triples, constructing a sentence tree, and employing graph attention networks. Entity extraction is treated as a sequence labeling problem, while relation extraction is approached as a classification task.
Similarly, \cite{huang2024ctikg} focuses on extracting knowledge triplets from CTI texts using off-the-shelf LLMs, but also incorporate an agent-based workflow.
Alternatively, the authors of \cite{ahmed2024cyberentrel} propose CyberEntRel, a model designed for the joint extraction of entities and relations from CTI data. The authors applied a \quotes{BIEOS} tagging scheme combined with an attention-based RoBERTa-BiGRU-CRF model for sequential tagging. The approach effectively extracts relation triples—two entities and their associated relationship—using a relation-matching technique.

Our approach stands out from the studies referenced in three significant ways.
Firstly, these studies do not prioritize facilitating threat knowledge sharing, where entity and relationship types are defined freely by the specific needs of an application rather than by a standard.
\system~is designed to generate threat knowledge that can be shared with any SOC tool or detection system that supports STIX, ensuring broader applicability and impact on practice.
Secondly, we break down the overall task into smaller, more manageable components, which allows models to perform with greater accuracy. This segmentation also permits human experts to participate actively in the process, providing guidance where necessary.
Lastly, we not only utilize off-the-shelf LLMs, which possess cybersecurity domain knowledge and the ability to follow instructions, but we also fine-tune them on real-world data to specifically tailor their performance to our tasks.


\textbf{STIX Report Generation:} Few work proposed to generate structured cyber threat information, complaint with the STIX standard, from unstructured text \cite{8616998,park2020automatic}.
In this regard, \cite{8616998} proposes a solution that asks the users to manually enter only STIX entities (without relationships) in a form. Then, the solution stores the entities in a database to allow analysts to make queries on the stored entities. 
\cite{park2020automatic} alternatively introduced a tool designed to convert Android malware analysis files (like logs) from the Malware Attribute Enumeration and Characterization (MAEC) format to the STIX format. 
This tool employs a dual-strategy for conversion: elements in the MAEC file that are amenable to standard automated analysis are identified for automatic conversion, while those needing manual analyst intervention are marked for separate processing.
Our evaluation of the tool in \cite{8616998} revealed that it fails to include key entities such as malware and tools, and crucially, it does not automatically handle relationships, requiring manual input. Conversely, the study in \cite{park2020automatic} focuses on converting Android malware analysis logs first to MAEC and then to STIX. Consequently, analysts are still required to manually review malware analysis logs to fill in any missing information.

\section{Conclusion}
\label{sec:conclusion}
In this paper, we introduced \system, a semi-automated tool to help practitioners create STIX reports by processing threat analysis reports. Our approach involves fine-tuning LLMs to extract entities and relationships as defined by the STIX standard. We detailed the challenges when using LLMs to process threat analysis reports, identified a series of tasks, and designed appropriate prompting strategies to effectively utilize the power of LLMs to generate STIX reports. We compiled an extensive dataset meticulously annotated with ground truth information following a rigorous process. Our evaluation of the dataset shows that \system~achieves high accuracy in terms of all tasks, and could help security analysts quickly summarize critical information of cyber threats into STIX reports. As there is limited availability of such datasets in the threat intelligence domain, we made our collected datasets available to the research community, which is another contribution of this work. For future work, \system~produces knowledge graphs that can be stored in a graph database. This will allow users to run queries on their graphs and discover hidden patterns especially that cross-reports relations can be modeled.


\bibliographystyle{IEEEtranS}

\appendices

\section{Impact of Hyperparameters on Performance}
\label{subsec:parameters_effect}

We investigated the effects of varying learning rates, temperature settings, and top-p values on the accuracy of the fine-tuned models.

\begin{figure}[!ht]
    \centering
    \scalebox{0.23}{\includegraphics{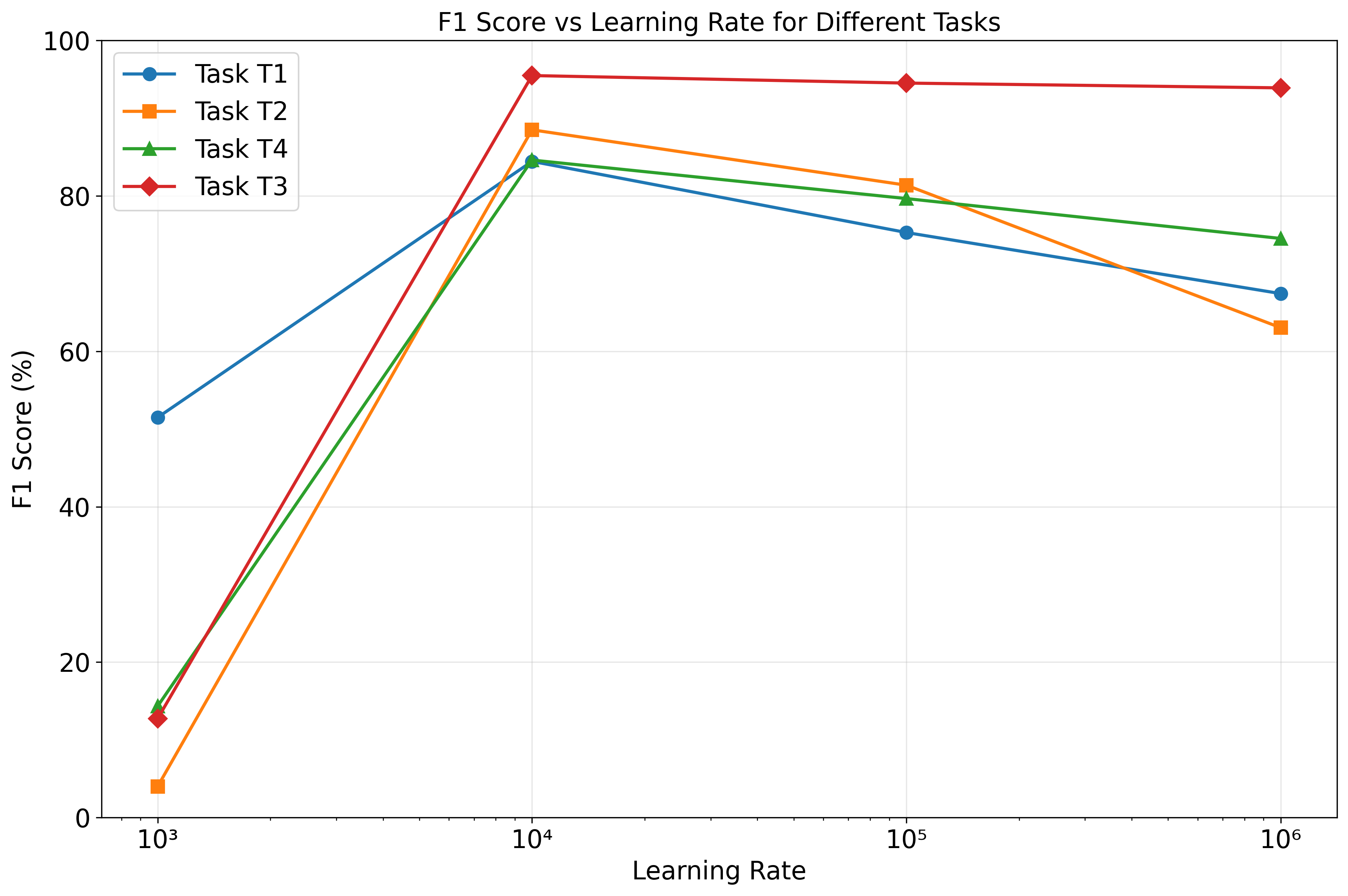}}
    \caption{Impact of learning rates (log scale) on F1-Score for tasks T1-T4 using the AZERG-MixTask model. A learning rate of $10^{-4}$ yielded the best overall performance.} 
    \label{fig:effect_of_learning_rates}
\end{figure}

\textbf{Changing learning rate.} 
The learning rate is a crucial parameter that controls the integration of newly acquired information with previously learned knowledge in machine learning models. Higher learning rate values may cause the model to overwrite previously encoded knowledge due to excessively large updates, potentially leading to poor generalization. Conversely, excessively low learning rates can impede the assimilation of new information, resulting in slow learning progress and the risk of underfitting. An optimal learning rate balances the absorption of new data with the retention of existing knowledge to ensure effective model training.
In our experiments, we fine-tuned the AZERG-MixTask model using a range of learning rates in  \{$10^{-6}$,  $10^{-5}$,  $10^{-4}$,  $10^{-3}$\}. 
The F1-scores for each task are shown in Fig. \ref{fig:effect_of_learning_rates}. 
Based on these results, a learning rate of  $10^{-4}$ has been determined to be optimal for the continual fine-tuning of the mistralai/Mistral-7B-Instruct-v0.3 model across all tasks.

\textbf{Changing sampling parameters.}
The temperature and top-p parameters shape the distribution of token probabilities during decoding. Higher temperature values flatten this distribution, increasing the selection of less likely tokens and adding randomness to the generated text. In contrast, lower temperatures favor more probable tokens, resulting in predictable outcomes. The top-p parameter controls the range of token selection. Higher top-p values include a wider array of tokens up to a certain cumulative probability, enhancing variability. Lower top-p values restrict this to the most likely tokens, thus producing more predictable text.
To select the optimal temperature and top-p values, we performed a grid search over a range of temperatures (0.0, 0.4, 0.7, 1.0, and 1.5) and top-p values (0.1, 0.4, 0.7, 0.95, and 1.0). 
Our search reveals that the optimal values for temperature and top-p are 0.7 and 0.1, respectively.

\begin{figure}[]
    \centering
    \includegraphics[width=3.0in]{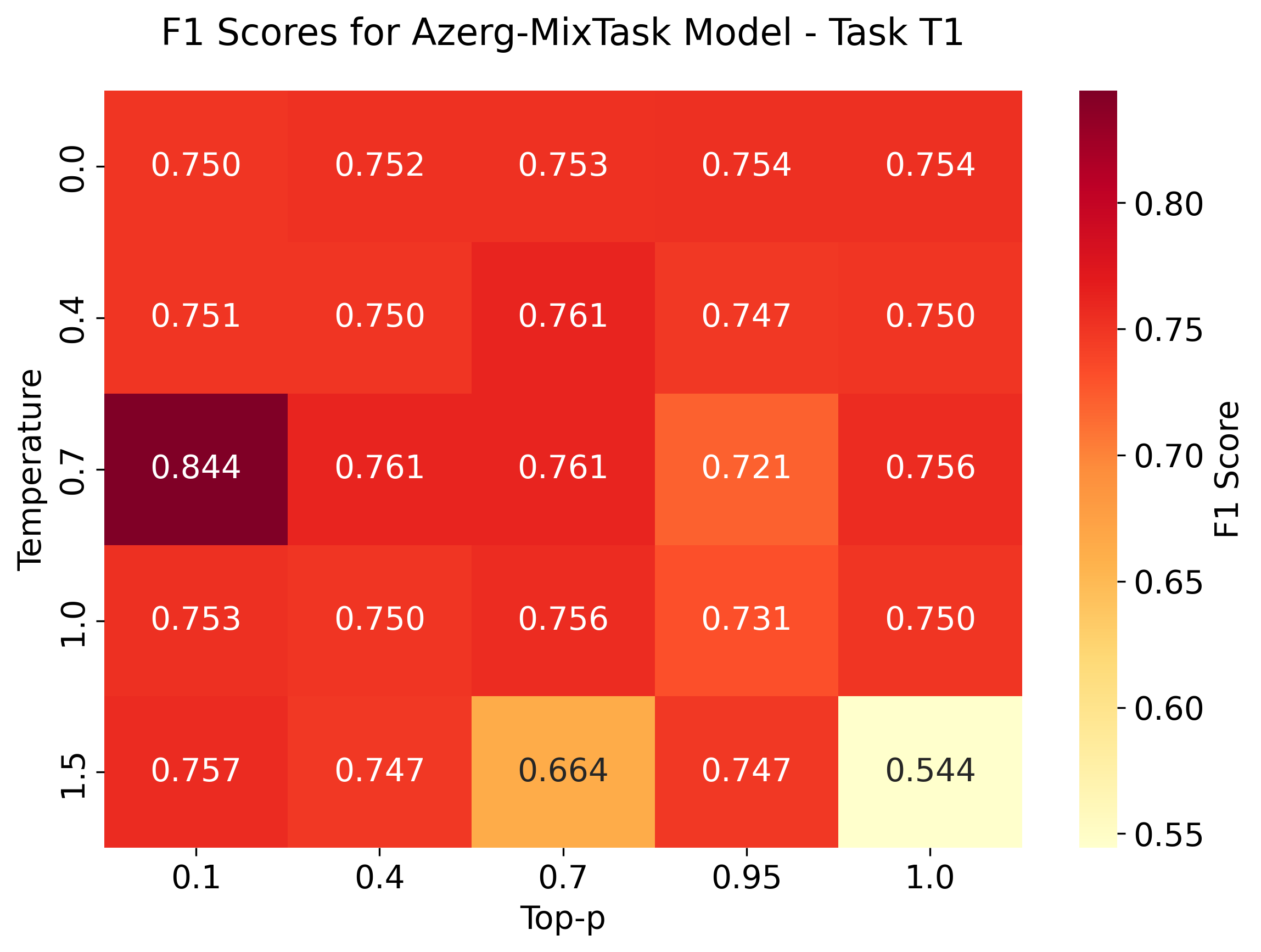}
    \caption{F1-score heatmap for T1.} 
    \label{fig:f1_score_heatmap_T1}
\end{figure}

\begin{figure}[]
    \centering
    \includegraphics[width=3.0in]{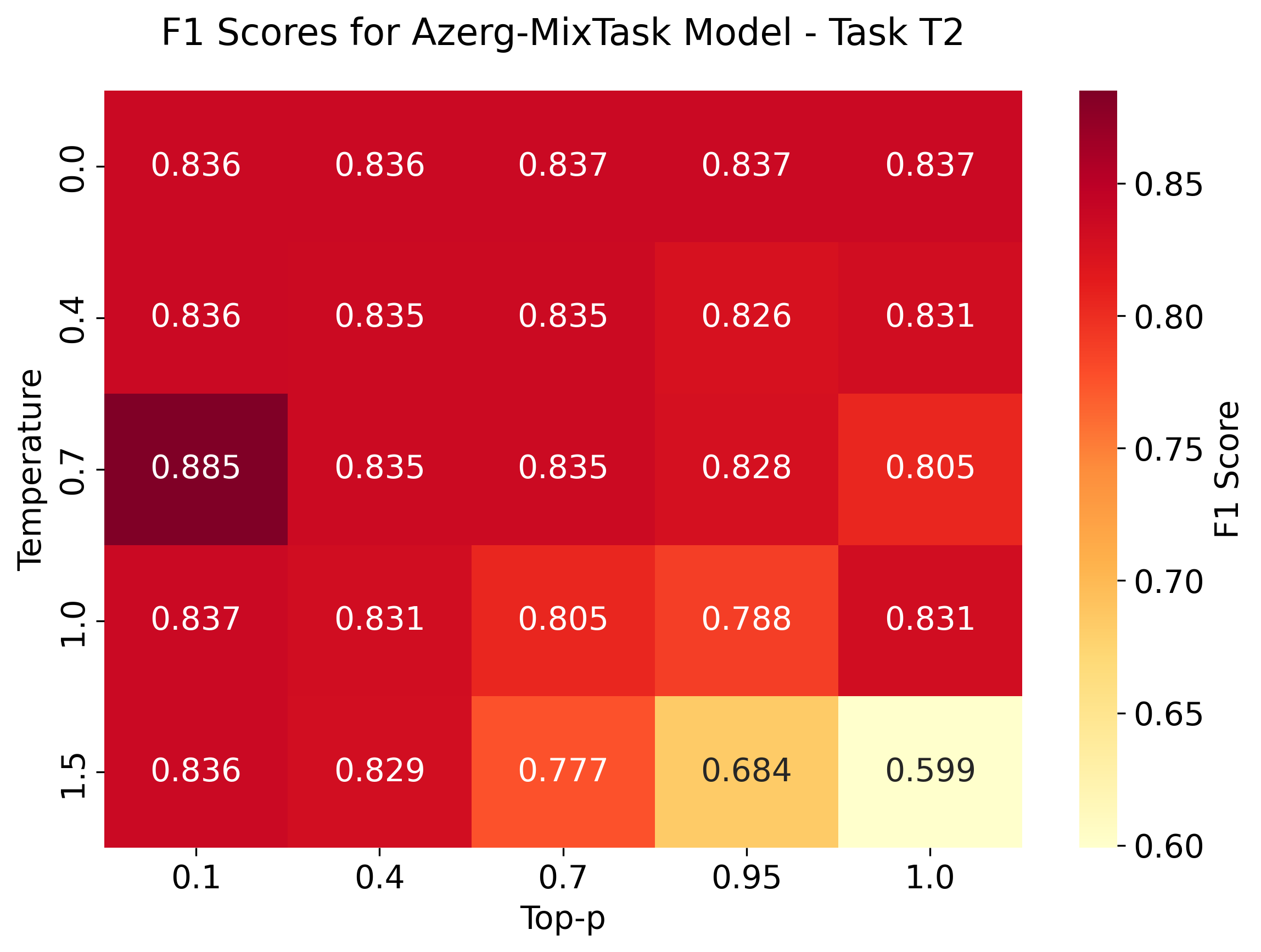}
    \caption{F1-score heatmap for T2.} 
    \label{fig:f1_score_heatmap_T2}
\end{figure}

\begin{figure}[]
    \centering
    \includegraphics[width=3.0in]{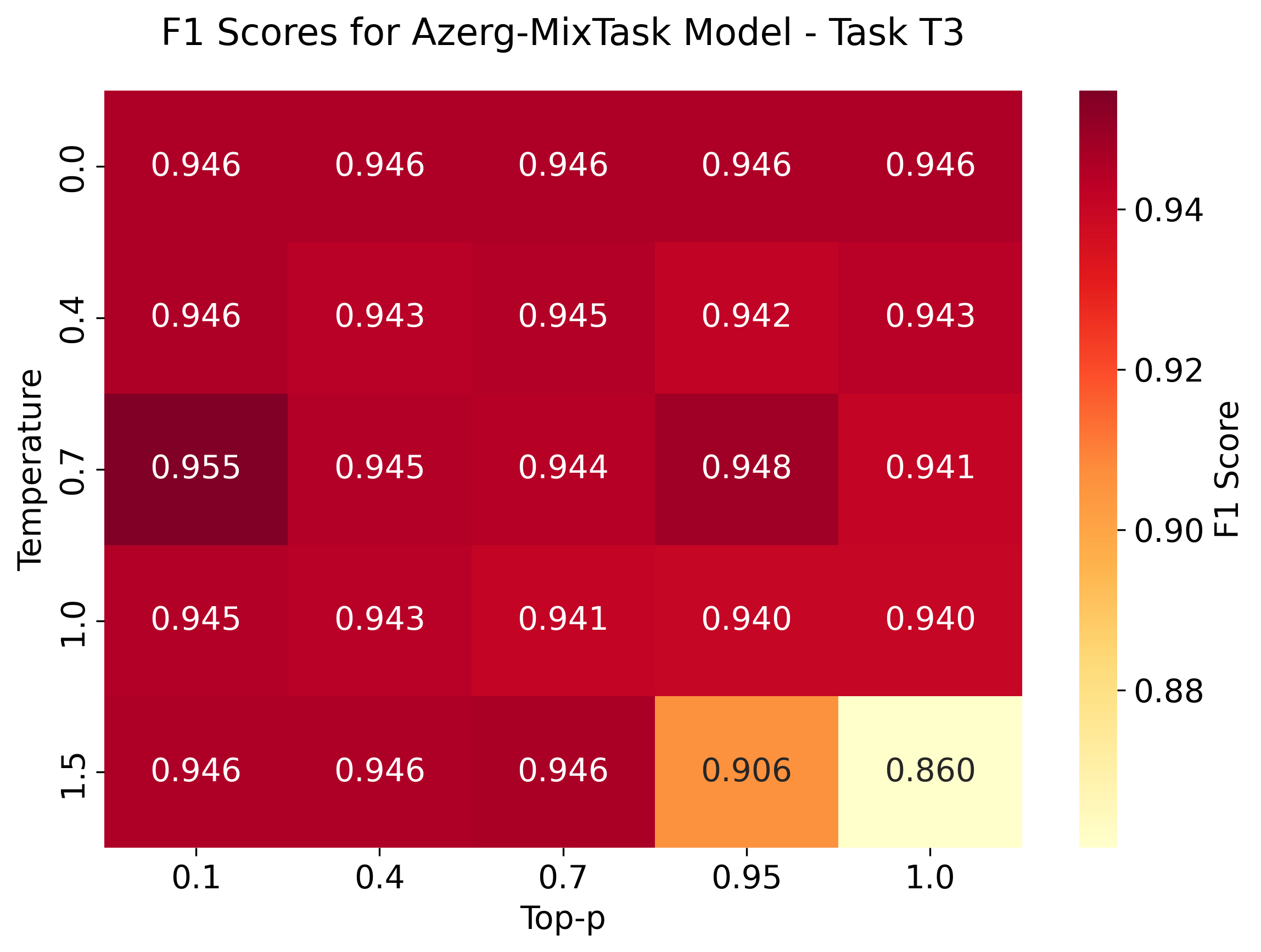}
    \caption{F1-score heatmap for T3.} 
    \label{fig:f1_score_heatmap_T3}
\end{figure}

\begin{figure}[]
    \centering
    \includegraphics[width=3.0in]{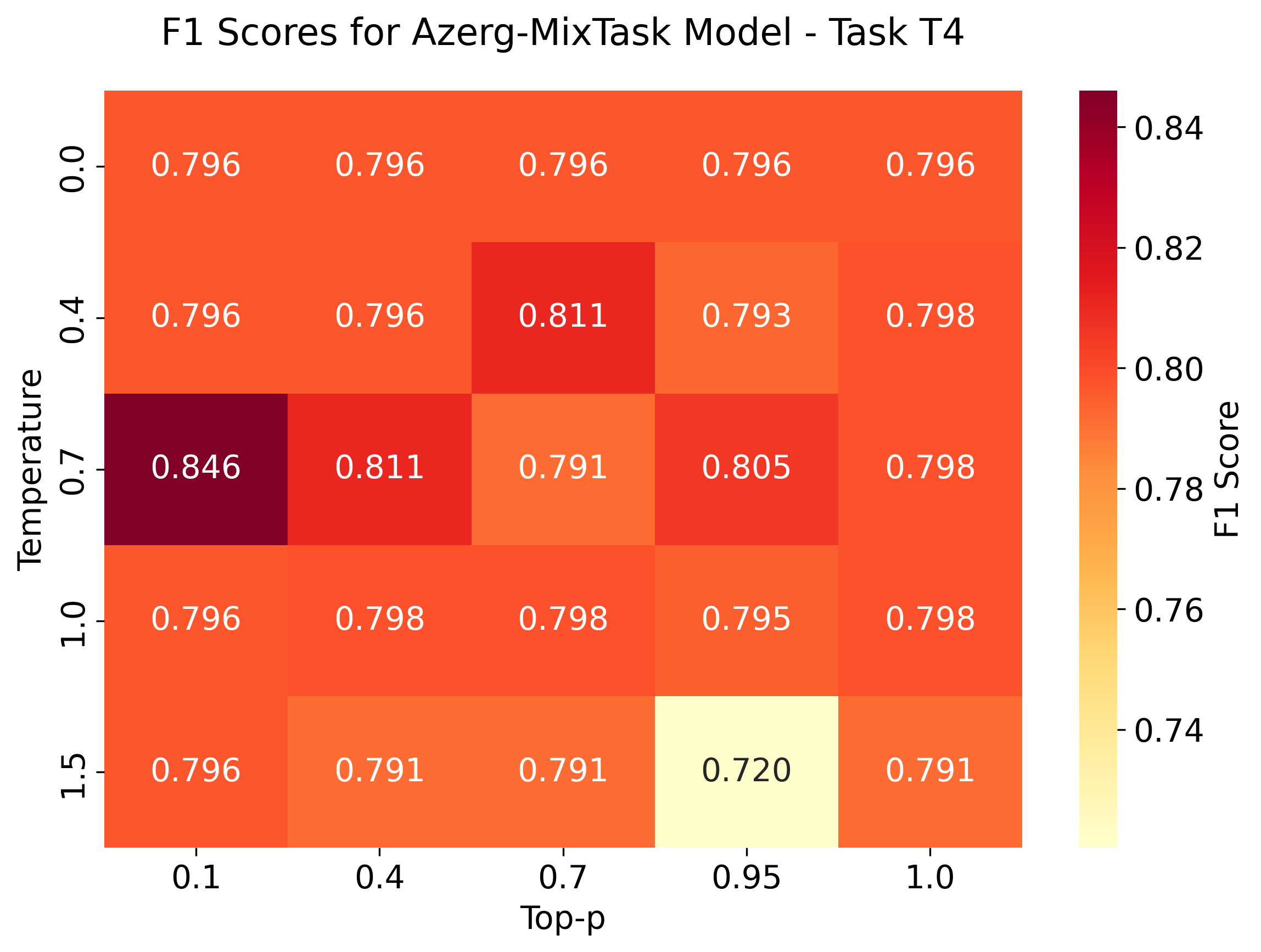}
    \caption{F1-score heatmap for T4.} 
    \label{fig:f1_score_heatmap_T4}
\end{figure}

We present the results of our grid search for the optimal setting for the temperature and top-p values. Fig. \ref{fig:f1_score_heatmap_T1}, Fig. \ref{fig:f1_score_heatmap_T2}, Fig. \ref{fig:f1_score_heatmap_T3}, and Fig. \ref{fig:f1_score_heatmap_T4} represent the F1-score heatmaps of T1, T2, T3, and T4, respectively.

We observe that lower temperatures (0.0 and 0.7) generally yield better F1-scores across most tasks, particularly for T2, where the F1-score is consistently high. 
As the temperature increases to 1.0, the performance very slightly decreases for all tasks, and beyond that there is a marked drop in F1-scores, indicating that higher temperatures negatively impact the model's predictive accuracy. In terms of top-p values, a top-p of 0.1 produces the best overall performance, with consistently high F1-scores across all tasks. 
As top-p increases, the F1-scores gradually decrease, with the lowest performance observed at a top-p of 1.0. This suggests that a more conservative sampling strategy (lower top-p) is more effective for this model in maintaining high accuracy across different tasks.

We note that T1 has a higher sensitivity to sampling parameters. This is due to the nature of the task that asks the model to extract all entities from a text. It does not limit the choices for the model as per the other tasks that give precise choices for the model to select from.
To this end, the optimal parameter pair is temperature 0.7 and top-p 0.1.

\section{Prompt Templates}
\label{apdx:prompt_templates}

In this section, we present the prompts we used for our fine-tuning and inferences stages. We note that the same prompts were used to evaluate the state-of-the-art language models.

\begin{figure}[!ht]
  \centering
    \llmprompt{
    \# Instruction: \\
    You are a helpful threat intelligence analyst. Your task is to extract all STIX entities mentioned in the input. To help you, here is a list of the possible STIX entity types. \\
    STIX entity types: \\
    - ATTACK\_PATTERN: A type of TTP that describes ways that adversaries attempt to compromise targets. (e.g., T1051, T1548.001, etc.) \\
    - CAMPAIGN: A grouping of adversarial behaviors that describes a set of malicious activities or attacks (sometimes called waves) that occur over a period of time against a specific set of targets. \\
    - [...] \\
    Answer in the following format: <entities>LIST OF IDENTIFIED ENTITIES SEPARATED BY PIPE | (e.g., Ent1|Ent2|...|Entn)</entities> \\
    \# Input: \\
    - Text Passage: [INPUT TEXT] \\
    \# Response:
    }
  \caption{The employed prompt for T1.}
  \label{prmt:t1_prompt}
\end{figure}

\begin{figure}[!ht]
  \centering
    \llmprompt{
    \# Instruction: \\
    You are a helpful threat intelligence analyst. Your task is to assign a STIX entity type to the given Entity in the input. To help you, here is a list of the possible STIX entity types. 
    [STIX ENTITY TYPES]
    Choose STIX ENTITY TYPE from list of possible answers: ["ATTACK\_PATTERN", "CAMPAIGN", "COURSE\_OF\_ACTION", "IDENTITY", "INDICATOR", "INFRASTRUCTURE", "LOCATION", "MALWARE", "THREAT\_ACTOR", "TOOL", "VULNERABILITY"].Answer in the following format: <entity\_type> ONE OF STIX ENTITY TYPES </entity\_type> \\
    \# Input: \\
    - Entity: [TARGET ENTITY] \\
    - Text Passage: [INPUT TEXT] \\
    \# Response:
    }
  \caption{The employed prompt for T2.}
  \label{prmt:t2_prompt}
\end{figure}

\begin{figure}[!ht]
  \centering
    \llmprompt{
    \# Instruction: \\
    You are a helpful threat intelligence analyst. Your task is to identify if the source entity and the target entity in the provided text passage are semantically related. To help you, we provide all the possible relationship labels between the source and target entities. If any label applies to the relationship, answer YES. Otherwise, answer NO.Answer in the following format: <related>YES or NO</related>\\
    \# Input: \\
    - Source Entity: [SOURCE ENTITY (ENTITY TYPE)] \\
    - Target Entity: [TARGET ENTITY (ENTITY TYPE)] \\
    - Possible Relationship Labels: [STIX RELATIONSHIP LABELS BETWEEN SOURCE AND TARGET ENTITIES]
    - Text Passage: [INPUT TEXT] \\
    \# Response:
    }
  \caption{The employed prompt for T3.}
  \label{prmt:t3_prompt}
\end{figure}

\begin{figure}[!ht]
  \centering
    \llmprompt{
    \# Instruction: \\
    You are a helpful threat intelligence analyst. Your task is to identify the label of the relationship between the source entity and the target entity in the provided text passage. To help you, we provide all the possible relationship labels between the source and target entities.Answer in the following format: <label>Your chosen label</label>\\
    \# Input: \\
    - Source Entity: [SOURCE ENTITY] \\
    - Target Entity: [TARGET ENTITY] \\
    - Possible Relationship Labels: [STIX RELATIONSHIP LABELS BETWEEN SOURCE AND TARGET ENTITIES] \\
    - Text Passage: [INPUT TEXT] \\
    \# Response:
    }
  \caption{The employed prompt for T4.}
  \label{prmt:t4_prompt}
\end{figure}

\end{document}